\documentclass[fleqn,usenatbib]{mnras}

\usepackage{newtxtext,newtxmath}
\usepackage[T1]{fontenc}
\usepackage{ae,aecompl}
\usepackage{graphicx}
\graphicspath{{Figures/}}
\usepackage{tabularx}
\usepackage{amsmath}
\usepackage{amssymb}
\usepackage{gensymb}
\usepackage{wasysym}
\usepackage{soul}
\usepackage{color}
\usepackage{times}

\newcolumntype{C}{>{\centering\arraybackslash}X}
\newcolumntype{L}{>{\raggedright\arraybackslash}X}
\newcolumntype{R}{>{\raggedleft\arraybackslash}X}
\newcommand*\diff{\mathop{}\!\mathrm{d}}

\defcitealias{McKinnon2017}{McK17}

\title[Radiative AGN feedback]{Radiative AGN feedback on a moving mesh: the impact of the galactic disc and dust physics on outflow properties}

\author[D. J. Barnes et al.]{David J. Barnes$^{1}$\thanks{E-mail: djbarnes@mit.edu},
Rahul Kannan$^{2}$\thanks{Einstein Fellow},
Mark Vogelsberger$^{1}$
and Federico Marinacci$^{1,2}$
\\
$^1${Department of Physics, Kavli Institute for Astrophysics and Space Research, Massachusetts Institute of Technology, Cambridge, MA 02139, USA}\\
$^2${Harvard--Smithsonian Center for Astrophysics, 60 Garden Street, Cambridge, MA 02138}\\
}

\date{Accepted XXX. Received YYY; in original form ZZZ}

\pubyear{2018}

\begin{document}
\label{firstpage}
\pagerange{\pageref{firstpage}--\pageref{lastpage}}
\maketitle

\begin{abstract}
Feedback from accreting supermassive black holes, active galactic nuclei (AGN), is now a cornerstone of galaxy formation models. In this work, we present radiation-hydrodynamic simulations of radiative AGN feedback using the novel \textsc{Arepo-RT} code. A central black hole emits radiation at a constant luminosity and drives an outflow via radiation pressure on dust grains. Utilising an isolated NFW halo we validate our setup in the single and multi-scattering regimes, with the simulated shock front propagation in excellent agreement with the expected analytic result. For a spherically symmetric NFW halo, an examination of the simulated outflow properties generated by radiative feedback demonstrates that they are lower than typically observed at a fixed AGN luminosity, regardless of the collimation of the radiation. We then explore the impact of a central disc galaxy and the assumed dust model on the outflow properties. The contraction of the halo during the galaxy's formation and modelling the production of dust grains results in a factor $100$ increase in the halo's optical depth. Radiation is then able to couple momentum more efficiently to the gas, driving a stronger shock and producing a mass-loaded $\sim10^{3}\,\mathrm{M}_{\astrosun}\,\mathrm{yr}^{-1}$ outflow with a velocity of $\sim2000\,\mathrm{km}\,\mathrm{s}^{-1}$, in agreement with observations. However, the inclusion of dust destruction mechanisms, like thermal sputtering, leads to the rapid destruction of dust grains within the outflow, reducing its properties below typically observed values. We conclude that radiative AGN feedback can drive outflows, but a thorough numerical and physical treatment is required to assess its true impact.
\end{abstract}

\section{Introduction}
\label{sec:intro}
Supermassive black holes (BHs) and their associated feedback are now a cornerstone upon which theories of galaxy and structure formation are built. When these BHs accrete material they are classified as active galactic nuclei (AGN) and release energy and momentum into the surrounding interstellar and intergalactic medium, profoundly influencing their host galaxy and halo. This creates a negative feedback loop that acts to limit the growth of BHs \citep[e.g.][]{Haehnelt1998,Wyithe2003,DiMatteo2005}.

AGN feedback is thought to be the mechanism that establishes the observed scaling relations between a galaxy's central BH and its large-scale properties, such as mass, luminosity and velocity dispersion \citep[e.g.][]{Gultekin2009,Heckman2011,Beifiori2012,McConnell2013}. Feedback is thought to regulate the baryonic content of the BH's host halo, ejecting gas, suppressing star formation and producing the observed population of quiescent massive galaxies \citep[e.g.][]{Scannapieco2004,Bower2006}. This process appears to be active even at $z=4$ \citep{Muzzin2013,Belli2015,Glazebrook2017}. On larger scales, AGN in galaxy groups and clusters are thought to prevent the onset of strong cooling flows \citep[e.g.][]{Peterson2003}, generate the observed cavities in the X-ray emission of the hot gas \citep[e.g.][]{Sanders2016} and play a significant role in the distribution of metals into the surrounding environment \citep[e.g.][]{Kirkpatrick2015,Mernier2017,Biffi2017,Vogelsberger2018}.

These observations are complemented by numerous detections of gas outflowing at high velocity. At parsec scales, X-ray absorption lines reveal the presence of winds with speeds up to $1000\,\mathrm{km}\,\mathrm{s}^{-1}$ \citep[e.g.][]{Blustin2005,McKernan2007,Tombesi2010,Reeves2013,King2015}. On kiloparsec scales, quasar and radio-loud AGN dominated spectra reveal the presence of mass-loaded, $\sim10^{8}\,\mathrm{M}_{\astrosun}$, multiphase outflows with velocities up to $3000\,\mathrm{km}\,\mathrm{s}^{-1}$ \citep[e.g.][]{Sturm2011,Genzel2014,Zakamska2014,Tombesi2015,Zakamska2016,Bischetti2017}. For a given AGN luminosity, $L$, the cold component of these outflows are characterized by large momentum fluxes $(>10L/c)$, high kinetic powers $(\sim0.05L)$ and high outflow rates $(\sim10^{3}\,\mathrm{M}_{\astrosun}\,\mathrm{yr}^{-1})$ \citep[e.g.][]{Zakamska2014,Cicone2015,Zakamska2016,Fiore2017}, where $c$ is the speed of light. These outflows are attributed to AGN because supernovae-driven winds are only capable of generating outflow velocities of at most $\sim600\,\mathrm{km}\,\mathrm{s}^{-1}$ \citep{Martin2005,SharmaNath2013}. Combined with observations that show the baryonic content of haloes is lower than the naive expectation \citep[e.g.][]{Shull2012}, the observed outflows are interpreted as AGN feedback in action and add further weight to the importance of supermassive BHs and AGN feedback in theoretical models of galaxy and structure formation.

In order to reproduce this wealth of observational evidence virtually all theories of galaxy formation require BHs and their associated feedback. AGN feedback is now ubiquitous in analytic \citep{SilkRees1998,Haehnelt1998}, semi-analytic \citep{Bower2006,Croton2006,Henriques2015} and hydrodynamical \citep{Springel2005,Sijacki2007,DiMatteo2008,BoothSchaye2009,Fabjan2010,Dubois2012,Vogelsberger2013,Volonteri2016,McCarthy2017} models of galaxy and structure formation. Recently, these models have begun to reproduce cosmologically representative volumes of the Universe with galaxies whose integrated properties are well matched to observations \citep[e.g.][]{Vnat2014,Vogelsberger2014,Schaye2015,Dave2017,Springel2018}. 
 
However, hydrodynamical simulations are prevented from \textit{ab initio} treatments of BHs and their associated feedback because of limited numerical resolution and a poor understanding of the relevant physical processes. This forces them to adopt a subgrid approach when modelling BHs and AGN feedback. In cosmological simulations BHs are placed in collapsed haloes as supermassive seeds that can grow by mergers and accrete gas via some variant of a Bondi-Hoyle model \citep[e.g.][]{DiMatteo2005,BoothSchaye2009,Rosas-Guevara2015,Steinborn2015}. AGN feedback is modelled by injecting energy and/or momentum into the gas surrounding the BH, with the ``mode'' of feedback commonly varied depending on the accretion rate \citep[e.g.][]{Sijacki2007,Weinberger2017a}. With the continued improvement in simulations, in terms of numerical resolution and galaxy formation models, it is becoming clear that current models of AGN feedback are no longer sufficient. In high-resolution cosmological simulations the failure of AGN feedback models leads to a range of issues. First, feedback does not eject enough material from the haloes of high-redshift progenitors of massive galaxies, resulting in stellar masses a factor of $2$ higher observed at fixed halo mass \citep{Bahe2017}. The gas fractions of massive haloes are also too high \citep{Barnes2017b,Henden2018,Tremmel2018,Barnes2018c}. Converting BH accretion rates to expected luminosity demonstrates that current models do not agree with the observed quasar luminosity function \citep{Rosas-Guevara2016,Weinberger2018}, especially at high redshift. Finally, current models yield an AGN duty cycle that is strongly driven by numerical considerations, which results in violent feedback events \citep{Barnes2017b,Barnes2018c} that are at odds with recent observations \citep{Hogan2017}. Physical processes missing from most simulations possibly play an important role in regulating AGN feedback \citep[e.g.][]{Kannan2017,Barnes2018b}, however, they do not alleviate all of the problems with current models.

A major issue with subgrid models of AGN feedback is that the mechanism for coupling the rest-mass energy of accreted material not swallowed by the BH to the surrounding medium, and the efficiency of this processes, is poorly constrained. This leads to most simulations simply calibrating the coupling efficiency to reproduce chosen observable relations \citep[e.g.][]{Vogelsberger2014,Schaye2015,McCarthy2017,Weinberger2017a}. The downside of this approach is that BH feedback is an ad-hoc scheme with a large number of free parameters that require fine-tuning. These parameters often require retuning with changes in numerical resolution, making the predictive power of such approaches limited at best. Therefore, to better understand AGN feedback it is critical to develop theoretical models for mechanisms that couple AGN feedback to the surrounding medium and to test these models in realistic setups. One mechanism is the launching of a fast nuclear wind from the accretion disc that interacts violently with the ISM and drives a blast wave outwards. Such models are able to account for the large momentum boosts and high kinetic luminosities observed \citep{King2003,FaucherGiguere2012,Wagner2013,Costa2014,Hopkins2016}. Models where radiation heats the ambient medium via Compton heating have also been shown to be effective at regulating accretion flows and star formation in galactic nuclei \citep{Hambrick2011,Kim2011}. Finally, radiation can couple directly to the ISM on large scales via radiation pressure on dust \citep{Fabian1999,Murray2005,Chattopadhyay2012,Novak2012,Thompson2015}. For sufficiently large optical depths, infrared (IR) photons scatter multiple times and the radiation is able to drive outflows whose global energetics are a reasonable match to observed outflows \citep{Ishibashi2018}.

The impact of radiation on structure formation is only just beginning to be understood \citep{Hopkins2012,Stinson2013,Aumer2013,Kannan2014a,Kannan2014b,Ceverino2014,Bieri2017,Cielo2017,Costa2018a,Costa2018b}, due to the computational challenges of the number of sources present in a cosmological simulation and the speed at which photons propagate through the simulation domain. In this paper, we explore radiation pressure on dust as a channel of AGN feedback using the recently implemented \textsc{Arepo-RT} radiative transfer (RT) scheme \citep{Kannan2018} within the \textsc{Arepo} moving-mesh code \citep{Springel2010}. In this initial work we focus on idealized, isolated haloes with Navarro-Frenk-White (NFW) profiles \citep{Navarro1997} to demonstrate that the implementation reproduces the results predicted by analytic models \citep{Murray2005,Thompson2015,Ishibashi2015,Ishibashi2016,Ishibashi2018}. In addition, we examine the properties of outflows driven by radiation pressure on dust and how these properties change with the conical focusing of the injected radiation. We then go beyond the simplified analytic setup to explore how a central disc galaxy and modelling the formation of dust, rather than assuming its composition, impact the outflow properties.

The paper is organized as follows. In Section \ref{sec:meth} we describe our numerical method, which includes the radiative transfer scheme, the isolated NFW setup, the treatment of the black hole, the dust physics assumed, the detection of the shock front and outflowing material and the analytic solution we compare against. We compare our simulations to expected analytic result in Section \ref{sec:sim_anlyt}, exploring different luminosities, resolutions, speed of light approximations and the impact of reprocessing UV radiation into IR radiation. In Section \ref{sec:outprops} we examine the properties of the outflows produced, the impact of collimating the radiation output and compare to observations. We then explore the properties of the disc galaxy that forms if the NFW halo is allowed to cool and how its presence and modelling dust production impact outflow properties in Section \ref{sec:dustphys}. In Section \ref{sec:lims} we discuss some of the caveats of this work before presenting our conclusions in Section \ref{sec:concs}.

\section{Numerical method}
\label{sec:meth}
In this section we detail our numerical method. First, we briefly describe the radiative transfer implementation used in the radiation-hydrodynamic (RHD) simulations presented in this work. We then detail the galaxy formation, black hole and dust physics models used, including how photons are injected into the simulations and the chosen dust opacities. The method used to detect the shock produced by the radiative feedback and to determine what material is within the resulting outflow is then presented. We conclude by outlining the properties of the isolated NFW halo that provides the initial conditions for all simulations presented and the result expected for the analytic model of such an NFW halo.

\subsection{Radiative transfer scheme}
The simulations are performed using the \textsc{Arepo} code \citep{Springel2010}, where the relevant hydrodynamic equations are evolved using a moving-mesh finite volume scheme. We make use of the mesh regularization scheme presented in \citet{Vogelsberger2012} and the improved time integration and gradient estimations techniques described in \citet{Pakmor2016}. 

A detailed description of the moving-mesh radiative transfer scheme and its validation via idealized tests is given in \citet{Kannan2018}. Here we briefly describe the scheme. The zeroth and first-order moments of the radiative transfer equation are solved using the M1 closure relation for the Eddington tensor \citep{Levermore1984,Dubroca1999,Ripoll2001}. This makes the scheme completely local in nature and the computational cost independent of the number of sources of radiation. The upwind nature of the scheme is always ensured. The Riemann problem at the interface between cells is solved using Godunov's approach \citep{Godunov1959} and the setup uses the Rusanov flux function \citep{Rusanov1961}. Higher order accuracy is obtained by replacing the piecewise constant approximation of Godunov's scheme with a slope-limited piece-wise linear spatial extrapolation and a first-order prediction forward in time to obtain the states of the primitive variables on either side of the interface \citep{vanLeer1979}. The implementation is fully conservative and compatible with the individual time stepping scheme of \textsc{Arepo}. The higher order nature of the scheme has been demonstrated to produce very low numerical diffusion and improved convergence properties \citep{Kannan2018}.

The scheme couples ultra-violet (UV) radiation to a Hydrogen and Helium thermochemistry network, allowing the radiation to couple to the gas via photoionization, photoheating and the radiation pressure of ionizing photons. The radiation, both in the UV and IR bands, is also coupled to dust grains, which enables the transfer of energy and momentum between the radiation and the dust in both the single and multiscattering regimes. In this work the dust dynamics are not followed self-consistently, we assume that the gas and the dust are perfectly hydrodynamically coupled \citep{Murray2005}. To allow us to focus on the impact of the radiation pressure on the dust grains we neglect photoionization and photoheating of the gas throughout this work.

Due to the speed at which photons propagate through the simulation domain the time-step size is limited by the speed of light. To prevent the time-step from becoming prohibitively small we make use of the speed of light approximation throughout this work \citep{Gnedin2001}. Our fiducial value is $\tilde{c}=0.1c$. As shown in Section \ref{sec:sim_anlyt}, we verify that the presented results are converged with respect to the chosen speed of light approximation. The radiative transfer step can, in theory, be subcycled in the simulations, i.e. for every hydro time-step performed the radiative transfer step is performed many times. This can result in a significant reduction in the computational cost of the simulations. However, in this work we set the number of subcycles to $N_{\mathrm{sub}}=1$. This is done to ensure that the region around the black hole is sufficiently sampled by gas cells and has a very minor effect for those simulations where the outflows are spherically symmetric.

\subsection{Galaxy formation model}
For results presented in Sections \ref{sec:sim_anlyt} and \ref{sec:outprops} the gas in the RHD simulations is adiabatic, i.e. does not cool radiatively and does not form stars. However, when exploring the impact of the assumed dust physics in Section \ref{sec:dustphys} we use a more realistic setup: an isolated NFW halo with a central disc galaxy. The gas in the isolated NFW halo is simulated without radiation for a further $2.5\,\mathrm{Gyr}$, during which time we include the basic physical processes required for galaxy formation. The core components of our chosen galaxy formation model are described in detail in the literature \citep{Vogelsberger2013,Torrey2014} and we only briefly outline them here. Gas is allowed to cool radiatively, including both primordial and metal line cooling for nine chemical elements (H, He, C, N, O, Ne, Mg, Si, Fe). Stars are stochastically created in dense regions, with masses distributed according to a \citet{Chabrier2003} initial mass function. As stars evolve they return mass and metals to their neighbouring gas cells. The stellar lifetime function is taken from \citet{Portinari1998}, and the yields are adopted from \citet{Karakas2010} for AGB stars, from \citep{Thielemann2003} for SNe Ia and from \citet{Portinari1998} for SNe II. We note that the stars never inject photons into the simulation. We do not include the black hole formation, growth or feedback component of the model in these runs. The result of this run is an isolated NFW halo with a central disc galaxy, which is used as the initial conditions for some of the runs presented in Section \ref{sec:dustphys}. We note that our galaxy formation model is very similar to the IllustrisTNG model with the magnetic fields, supernovae winds and BH modelling removed.

\subsection{Black hole treatment and radiation injection}
For comparison purposes the modelling of the black hole is kept relatively simple throughout this work. We place a single black hole particle with a mass $m_{\mathrm{BH}}=10^8\,\mathrm{M}_{\astrosun}$ at the centre of the initial halo. During the relaxation, the formation of the disc galaxy (if required) and the RHD simulations the black hole is not fixed at the potential minimum, but we have confirmed that it remains within $0.1\,\mathrm{kpc}$ of the potential minimum at all times. The black hole does not accrete material from its surroundings or change in mass in any way during the simulations. In the RHD simulations, the black hole emits radiation at a constant bolometric luminosity in the UV band only, mimicking the analytic solution. At each time-step the black hole is active the number of photons released is given by its prescribed luminosity and the size of its current time-step, assuming a monochromatic radiation with an energy $E_{\gamma}=13.6\,\mathrm{eV}$. As we neglect photoionization and photoheating the energy of individual photons is academic. 

The radiation is injected into the nearest $256$ gas cells neighbouring the black hole. The number of neighbours is kept fixed across the different resolutions presented and we have confirmed that it has a negligible impact on the results presented. Each cell receives a fraction of the emitted photons based on the fraction of solid angle it covers relative to the total solid angle covered by the selected $256$ neighbours. We compute the solid angle via
\begin{equation}
 \Omega = \frac{1}{2}\left(1-\frac{1}{\sqrt{1+A\,/\,\mathrm{\pi}r^{2}}}\right)\:,
\end{equation}
where $A$ is the area of the gas cell. Therefore, cells that are closer to the black hole typically receive more photons than those at the edge of the injection region. We found that a solid-angle-weighted rather than mass-weighted injection significantly reduced the noise associated with the underlying cell distribution. When photons are injected into a cell the flux vector is explicitly set so that it points radial away from the black hole, which improves the collimation of the outflow for the simulations presented in Section \ref{sec:outprops}.

\subsection{Dust physics}
The response of the halo to radiative feedback and the properties of the outflow that develop depend strongly on the optical depth of the halo. This is governed by the assumptions made about the properties and distribution of dust grains in the halo. For UV radiation, previous analytic and numerical work have assumed an absorption coefficient of $10^{5}\,\mathrm{cm}^{2}\,\mathrm{g}^{-1}$, which combined with a Milky Way-\textit{like} dust-to-gas ratio $D=0.01$ \citep{LiDraine2001} yields a realistic UV dust opacity of $\kappa_{\mathrm{UV}}=10^{3}\,\mathrm{cm}^{2}\,\mathrm{g}^{-1}$. The caveat to this work is the strong assumption that the dust-to-gas ratio of $1$ per cent is true throughout the entire halo.

For IR radiation, the use of a halo with an NFW profile presents an issue. Due to its inner slope of $1/r$, an NFW profile converges to a maximum column density and it becomes very difficult to achieve a sufficiently high optical depth in the halo at $z=0$. This reduces the momentum boost seen by the inclusion of IR radiation to a negligible value. As previously demonstrated, even for very concentrated haloes the momentum boost due to multiply scattered IR radiation is still relatively small \citep[e.g.][]{Costa2018a}. Previous analytic and numerical work have circumvented this issue by using an unrealistically high IR dust opacity.

To enable comparisons to the expected analytic solution and previous work in Sections \ref{sec:sim_anlyt} and \ref{sec:outprops} we assume that the dust-to-gas ratio is fixed to $1$ per cent throughout the volume of the halo. In addition, we make use of an unrealistically high IR dust opacity when IR radiation is included. This yields dust opacities of $\kappa_{\mathrm{UV}}=10^{3}\,\mathrm{cm}^{2}\,\mathrm{g}^{-1}$ and $\kappa_{\mathrm{IR}}=10^{3}\,\mathrm{cm}^{2}\,\mathrm{g}^{-1}$ for UV and IR radiation, respectively. Given the chosen dust opacities the initial NFW halo has an optical depth $\tau=11.37$ when integrated out to $r_{200}$.

\renewcommand\arraystretch{1.2}
\begin{table*}
 \caption{Parameters of the radiation hydrodynamics simulations presented in this work. The columns are the simulation name, bolometric luminosity of the central black hole, target gas mass, gas gravitational softening, the dark matter mass for runs with a live halo, the dark matter gravitational softening, the opening angle of the radiation injection region, the inclusion of reprocessing of the UV radiation to IR radiation, the speed of light approximation, whether the galaxy formation model was included and the dust model assumed (either a constant dust-to-gas ratio, $D=0.01$, or the \citetalias{McKinnon2017} dust formation model), respectively.}
 \begin{tabularx}{\textwidth}{l c c c c c c c c c c}
 \hline
 Name & $L_{\mathrm{AGN}}$ & $m_{\mathrm{gas}}$ & $\epsilon_{\mathrm{gas}}$ & $m_{\mathrm{DM}}$ & $\epsilon_{\mathrm{DM}}$ & $\theta_{\mathrm{\gamma}}$ & IR & $\tilde{c}/c$ & Galaxy Formation & Dust Formation \\
  & $(\mathrm{erg}\,\mathrm{s}^{-1})$ & $(\mathrm{M}_{\astrosun})$ & $(\mathrm{pc})$ & $(\mathrm{M}_{\astrosun})$ & $(\mathrm{pc})$ & $(\degree)$ & reprocessing & & Model & Model \\
  \hline
  NFW\_UV46 & $10^{46}$ & $10^{4}$ & $83$ & Static & -- & $90$ & -- & 0.10 & -- & $D=0.01$ \\
  NFW\_UV47 & $10^{47}$ & $10^{4}$ & $83$ & Static & -- & $90$ & -- & 0.10 & -- & $D=0.01$ \\
  NFW\_UV47\_m1e5 & $10^{47}$ & $10^{5}$ & $177$ & Static & -- & $90$ & -- & 0.10 & -- & $D=0.01$ \\
  NFW\_UV47\_m1e6 & $10^{47}$ & $10^{6}$ & $378$ & Static & -- & $90$ & -- & 0.10 & -- & $D=0.01$ \\
  NFW\_UVIR46 & $10^{46}$ & $10^{4}$ & $83$ & Static & -- & $90$ & $\checkmark$ & 0.10 & -- & $D=0.01$ \\
  NFW\_UVIR47 & $10^{47}$ & $10^{4}$ & $83$ & Static & -- & $90$ & $\checkmark$ & 0.10 & -- & $D=0.01$ \\
  NFW\_UVIR47\_$\tilde{c}$0p01 & $10^{47}$ & $10^{4}$ & $83$ & Static & -- & $90$ & $\checkmark$ & 0.01 & -- & $D=0.01$ \\
  NFW\_UVIR47\_$\tilde{c}$1p00 & $10^{47}$ & $10^{4}$ & $83$ & Static & -- & $90$ & $\checkmark$ & 1.00 & -- & $D=0.01$ \\
  \hline
  NFW\_UV47\_$\theta$15 & $10^{47}$ & $10^{4}$ & $83$ & Static & -- & $15$ & --& 0.10 & -- & $D=0.01$ \\
  NFW\_UV47\_$\theta$30 & $10^{47}$ & $10^{4}$ & $83$ & Static & -- & $30$ & -- & 0.10 & -- & $D=0.01$ \\
  NFW\_UV47\_$\theta$45 & $10^{47}$ & $10^{4}$ & $83$ & Static & -- & $45$ & -- & 0.10 & -- & $D=0.01$ \\
  NFW\_UV47\_$\theta$60 & $10^{47}$ & $10^{4}$ & $83$ & Static & -- & $60$ & -- & 0.10 & -- & $D=0.01$ \\
  \hline
  Gal\_UV47 & $10^{47}$ & $10^{5}$ & $176$ & $10^{6}$ & $365$ & $90$ & -- & 0.10 & $\checkmark$ & $D=0.01$ \\
  Gal\_UV47\_McK17nd & $10^{47}$ & $10^{5}$ & $176$ & $10^{6}$ & $365$ & $90$ & -- & 0.10 & $\checkmark$ & \citetalias{McKinnon2017} -- no destruction \\
  Gal\_UV47\_McK17 & $10^{47}$ & $10^{5}$ & $176$ & $10^{6}$ & $365$ & $90$ & -- & 0.10 & $\checkmark$ & \citetalias{McKinnon2017} \\
  \hline
  \end{tabularx}
 \label{tab:runs}
\end{table*}
\renewcommand\arraystretch{1.0}

However, in Section \ref{sec:dustphys} we explore the impact of the assumptions made about the dust. We achieve this by modelling the production of dust grains during the formation of the disc galaxy initial conditions. Here we briefly review the dust formation model, but it is described in detail in \citet{McKinnon2016} and \citet[][henceforth \citetalias{McKinnon2017}]{McKinnon2017}. The mass of dust in each chemical species is tracked for each individual gas cell. The dust is assumed to be fully coupled to the gas and is essentially a passive scalar that is advected between gas cells when the the hydrodynamic equations are solved. As stars evolve, a fraction of the metal species returned to gas cells are assumed to condense into dust, excluding N and Ne. The efficiency with which metals condensate follows \citet{Dwek1998}, but with a minor modification to the oxygen coefficient to avoid more dust forming than the oxygen mass released by a star. Although there is considerable uncertainty in the condensation efficiencies, the dust produced has been shown to be insensitive to moderate changes in their values \citep{McKinnon2016}. Following the prescription of \citet{Dwek1998} and \citet{Hirashita1999} the dust mass of gas cell $i$, $M_{\mathrm{i,dust}}$, grows via collisions with gas-phase metals according to
\begin{equation}
 \frac{\diff M_{\mathrm{i,dust}}}{\diff t}=\left(1-\frac{M_{\mathrm{i,dust}}}{M_{\mathrm{i,metal}}}\right)\left(\frac{M_{\mathrm{i,dust}}}{\tau_{\mathrm{g}}}\right)\:,
\end{equation}
where $M_{\mathrm{i,metal}}$ is the gas-phase metal mass of the cell and the dust growth timescale, $\tau_{\mathrm{g}}$, is given by
\begin{equation}
 \tau_{\mathrm{g}}=\tau^{\mathrm{ref}}_{\mathrm{g}}\left(\frac{\rho^{\,\mathrm{ref}}}{\rho}\right)\left(\frac{T^{\,\mathrm{ref}}}{T}\right)^{1/2}\:,
\end{equation}
where $\rho$ and $T$ are the density and temperature of the gas cell, $\rho^{\,\mathrm{ref}}=1$ H atom $\mathrm{cm}^{-3}$ and $T^{\,\mathrm{ref}}=20\,\mathrm{K}$ are the reference density and temperature of molecular clouds and $\tau^{\mathrm{ref}}_{\mathrm{g}}$ is the overall normalization factor that is derived in detail in \citet{Hirashita2000}. In addition, the model accounts for the destruction of dust grains by shocks and thermal sputtering according to
\begin{equation}
 \frac{\diff M_{\mathrm{i,dust}}}{\diff t}=-\left(\frac{M_{\mathrm{i,dust}}}{\tau_{\mathrm{d}}}\right)-\left(\frac{M_{\mathrm{i,dust}}}{\tau_{\mathrm{sp}}/3}\right)\:,
\end{equation}
where the dust destruction timescale due to supernovae, $\tau_{\mathrm{d}}$, is given by
\begin{equation}
 \tau_{\mathrm{d}}=\frac{m_{\mathrm{gas}}}{\beta\eta M_{\mathrm{s}}(100)}\:,
\end{equation}
where $m_{\mathrm{gas}}$ is gas mass of a cell, $\beta$ is the efficiency with which grains are destroyed in SN shocks, $\eta$ is
the local Type II SN rate, and $M_{\mathrm{s}}(100)$ is the mass of gas shocked to at least $100\,\mathrm{km}\,\mathrm{s}^{-1}$ \citep{Dwek1980,Seab1983,McKee1989}. The dust destruction timescale due to thermal sputtering is given by equation 14 of \citet{Tsai1995}
\begin{equation}
 \tau_{\mathrm{sp}} = a\left|\frac{\diff a}{\diff t}\right|^{-1}\approx(0.17\,\text{Gyr})\left(\frac{a_{-1}}{\rho_{-27}} \right)\left[\left(\frac{T_0}{T} \right)^\omega+1\right],
\end{equation}
where $a$ is the dust grain size, $a_{-1}$ is the grain size in units of $0.1\,\mu\mathrm{m}$, $\rho_{-27}$ is the gas density in units of $10^{-27}\,\mathrm{g}\,\mathrm{cm}^{-3}$ (which corresponds to a number density of $n\approx6\times10^{-4}\,\mathrm{cm}^{-3}$), $T_{0}=2\times10^{6}\,\mathrm{K}$ is the temperature above which the sputtering rate is constant and $\omega=2.5$ controls the sputtering rate at low temperatures. The model does not track the grain size distribution \citep[e.g.][]{McKinnon2018}, but instead assumes a fixed grain radius $a=0.1\,\mathrm{\mu m}$. When modelling the formation of dust we still assume an absorption coefficient of $10^{5}\,\mathrm{cm}^{2}\,\mathrm{g}^{-1}$.

\subsection{Shock and outflow detection}
\label{sec:outflowdef}
Comparison with the analytic expectation and observations requires knowledge of the shock front location and a definition of outflowing material. The analytic solution assumes that all material is swept out and occupies a thin shell at the position of the shock. However, in numerical simulations the shock front has a finite width due to the fixed numerical resolution of the simulation, the finite mean free path of the UV radiation and thermal broadening of the shock front. We locate the position of the shock by first creating a halocentric radial density profile in $100$ logarithmically space bins in the range $0.1$ to $220\,\mathrm{kpc}$. We compute the steepest gradient of the density profile and then create a second radial density profile using $100$ linearly spaced bins in a $10\,\mathrm{kpc}$ region centred on the steepest gradient. The position of the maximum density gradient for the linearly spaced profile is then taken to be the position of the shock front. The use of the second linearly space density profile prevents the shock being found in the same logarithmically spaced bin for multiple simulation snapshots at late times. We quantify the uncertainty on the position of the shock by the region where the density gradient is greater than half of the maximum value. For the linearly spaced density profile we apply a Savitzky-Golay filter, with a window width of $2\,\mathrm{kpc}$, to remove high-frequency noise, which is mainly relevant for the lower numerical resolution convergence runs.

We determine whether or not a gas cell is part of the outflow via its halocentric radial velocity component. If a gas cell has a radial velocity greater than $50\,\mathrm{km}\,\mathrm{s}^{-1}$ then it contributes to the outflow properties at that snapshot. This limit avoids the inclusion of spurious material that has not yet been affected by the feedback but is instead orbiting in the outskirts of the halo. In reality, material that fails to achieve escape velocity will eventually fall back to the centre of the halo, however, we verified that cutting at $50\,\mathrm{km}\,\mathrm{s}^{-1}$ or the escape velocity of the halo leads to a negligible difference in the properties of the outflow.

\subsection{Initial conditions}
To enable comparisons to the known analytic solution we keep the numerical setup relatively simple. The initial conditions consist of an isolated NFW halo centered on the centre of the computational domain with a virial mass $M_{200}=10^{12}\,\mathrm{M}_{\astrosun}$, a virial radius $r_{200}=212.08\,\mathrm{kpc}$\footnote{$M_{200}$ and $r_{200}$ are defined as the mass enclosed within and the radius of a sphere, respectively, whose average density is $200$ times the critical density of the Universe at $z=0$.}, a concentration $C=r_{\mathrm{s}}/r_{200}=10$, where $r_{\mathrm{s}}$ is the scale radius, and a gas fraction $F_{\mathrm{gas}}=0.1$. The simulation domain is a periodic cube with a side length of $820\,\mathrm{kpc}$. The gas follows an NFW profile to $r_{200}$, at which point the profile is truncated and a background grid of gas is added with a minimum density of $6.0\times10^{31}\,\mathrm{g}\,\mathrm{cm}^{-3}$. Gas cells have a fiducial target mass $m_{\mathrm{gas}}=10^{4}\,\mathrm{M}_{\astrosun}$ and a softening length $\epsilon=83\,\mathrm{pc}$. In the adiabatic gas simulations of Section \ref{sec:sim_anlyt} and \ref{sec:outprops}, the dark matter is modelled as a static, non-periodic potential, negating any noise that may be introduced by a live dark matter halo. In Section \ref{sec:dustphys}, we use a live dark matter halo to capture the contraction of the halo due to the formation of the disc galaxy. The dark matter particles in these simulations have a mass $m_{\mathrm{DM}}=10^{6}\,\mathrm{M}_{\astrosun}$ and a softening length $\epsilon=365\,\mathrm{pc}$. To remove any artefacts from the initial NFW halo it is relaxed for $1\,\mathrm{Gyr}$ in an adiabatic hydrodynamic run, for both the static and live dark matter setups. We verify that the halocentric radial density profile between the convergence radius \citep{Power2003} and $r_{200}$ deviates from the analytic expectation by less than $2$ per cent in any radial bin throughout the relaxation run. This relaxed halo with a static dark matter potential provides the initial conditions for all simulations presented in \ref{sec:sim_anlyt} and \ref{sec:outprops}, and provides the initial conditions for the galaxy formation run that results in an NFW halo with a central disc galaxy. Table \ref{tab:runs} provides a list of simulations presented in this work, denoting the physical processes included and parameters used.

\subsection{Analytic solution}
To demonstrate that the numerical setup is performing as expected we require an analytic solution to compare with. As an AGN feedback channel, radiation pressure on dust has been well studied via analytical models \citep{Murray2005,Ishibashi2015,Thompson2015,Ishibashi2016,Ishibashi2018}. We follow the analytic model presented in \citet{Ishibashi2015} and \citet{Thompson2015}, also see \citet{Costa2018a}.

Consider a spherically symmetric NFW halo with a total mass profile $M(R)$ and a gas fraction $F_{\mathrm{gas}}$. Assume that the gas within the halo perfectly traces the underlying dark matter distribution. Well mixed within the gas is dust, at a fixed dust-to-gas ratio $D=0.01$, that is perfectly hydrodynamically coupled to the gas. Therefore, a gas shell of width $dr$ at a radial distance $r$ has an optical depth of $d\tau=\kappa\rho(r)dr$, where $\kappa$ is the opacity of the gas to the radiation. At the origin of the halo is a black hole that is emitting radiation with a constant luminosity $L$. The radiation will exert a pressure on the gas, via its dust content, driving an outflow and sweeping up all material into a thin shell of mass $M_{\mathrm{SH}}$ as it propagates from the centre of the halo. When the radiation has propagated a distance $r$ the momentum equation for  a shell of mass $M_{\mathrm{SH}}$ is given by
\begin{equation}
 \frac{d[M_{\mathrm{SH}}(r)\dot{r}]}{dt}=f[\tau(r)]\frac{L}{c}-\frac{GM_{\mathrm{SH}}(r)M(r)}{r^{2}}\:,
\end{equation}
where $G$ is Newton's gravitational constant, $f[\tau(r)]=(1-e^{-\tau_{\mathrm{UV}}})+\tau_{\mathrm{IR}}$ is the fraction of momentum transferred to the gas shell by the incident radiation \citep{Thompson2015}, and we have neglected the external pressure from the ambient  medium. The function $f(\tau)$ accounts for the optical depth of the gas for both the singly scattered UV photons, $\tau_{\,\mathrm{UV}}$, and the multiply scattered, trapped IR photons, $\tau_{\,\mathrm{IR}}$. Assuming the gas and dark matter trace each other $M_{\mathrm{SH}}=F_{\mathrm{gas}}M$ and the equation of motion simplifies to
\begin{equation}\label{eq:eom}
 \frac{d[F_{\mathrm{gas}}M(r)\dot{r}]}{dt}=f[\tau(r)]\frac{L}{c}-\frac{F_{\mathrm{gas}}GM^{2}(r)}{r^{2}}\:.
\end{equation}

\begin{figure*}
 \includegraphics[width=\textwidth]{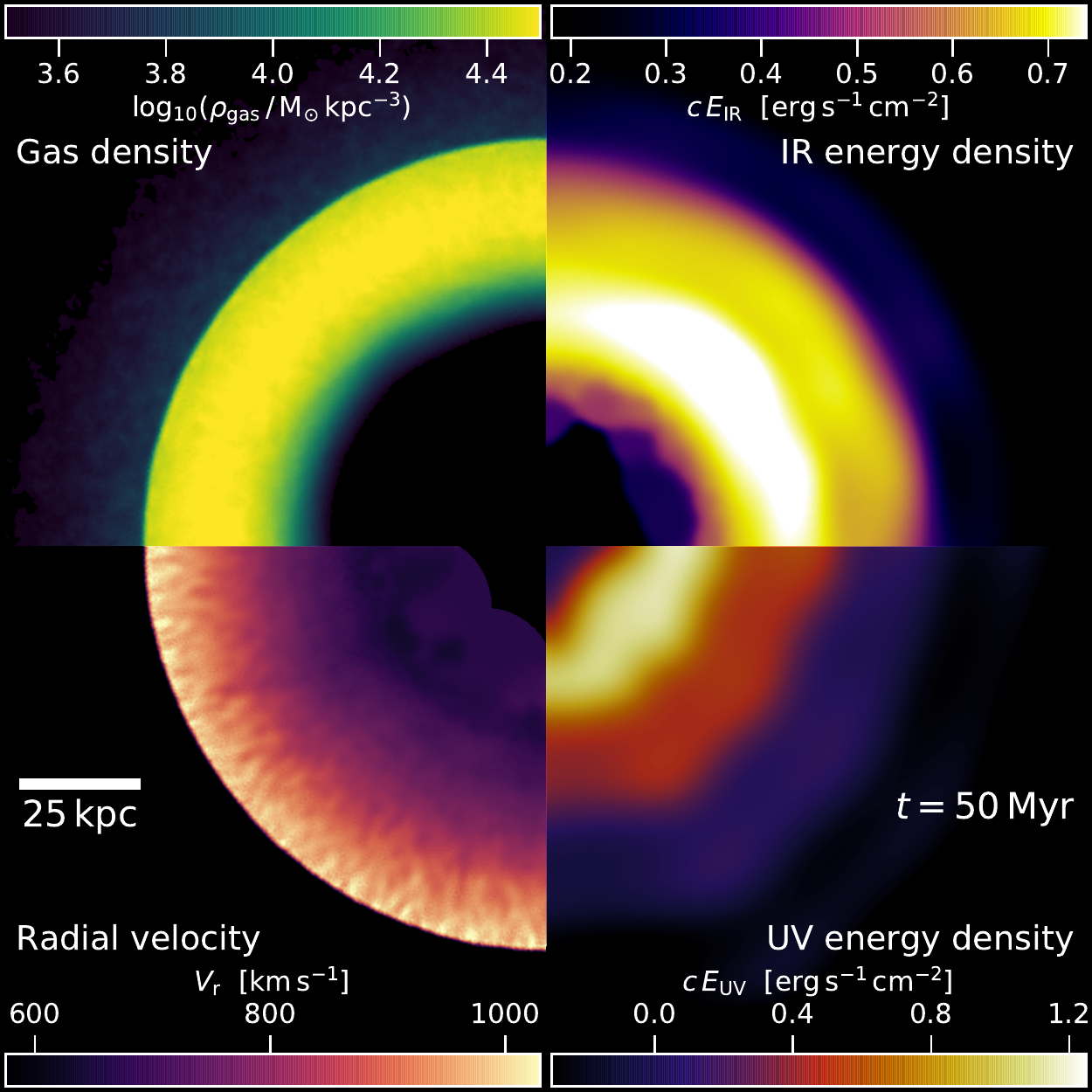}
 \caption{Slice of width $10\,\mathrm{kpc}$ through an RHD simulation with a central black hole isotropically emitting radiation at a constant luminosity $L_{\mathrm{AGN}}=10^{47}\,\mathrm{erg}\,\mathrm{s}^{-1}$ at $t=50\,\mathrm{Myr}$. Clockwise from the top left panel, quantities plotted are the gas density, IR energy density, UV energy density and radial velocity, respectively. The radiative feedback drives a shock front through the halo that propagates at $\sim1000\,\mathrm{km}\,\mathrm{s}^{-1}$. UV radiation is absorbed by dust grains, its energy density declining with radial distance from the black hole, and re-processed into IR radiation. Due to the chosen IR opacity of the dust, the IR photons are trapped behind the high-density shock front.}
 \label{fig:Outschm}
\end{figure*}

\noindent For an NFW profile, the gas mass enclosed within a given radius is
\begin{equation}
\begin{split}
 M(<r^\prime)&=F_{\mathrm{gas}}M_{200}\left[\frac{\ln(1+Cr^\prime)-Cr^\prime/(1+Cr^\prime)}{\ln(1+C)-C/(1+C)}\right] \\ & \equiv F_{\mathrm{gas}}M_{200}\left[\frac{g(Cr^\prime)}{g(C)}\right]\:,\\
\end{split}
\end{equation}
where we have used the dimensionless radius $r^\prime\equiv r/r_{200}$ and $g(r)\equiv\ln(1+r)-r/(1+r)$. The characteristic velocity of the halo is given by its virial velocity $v_{200}=\sqrt{GM_{200}/r_{200}}$, which in turn allows a characteristic time-scale $t_{200}=r_{200}/v_{200}$ for the halo to be established. Defining the dimensionless time variable $t^\prime=t/t_{200}$, equation (\ref{eq:eom}) can be written as
\begin{equation}\label{eq:eom2}
 \frac{d}{dt}\left[g(Cr^\prime)\frac{dr^\prime}{dt^\prime}\right]=f[\tau(r)]\frac{g(C)G}{F_{\mathrm{gas}}v^{4}_{200}}\frac{L}{c}-\frac{g^2(Cr^\prime)}{g(C)r^{\prime\,2}}\:.
\end{equation}
A thin shell at radius $r^\prime$ that has swept up all gas will have an optical depth
\begin{equation}
 \tau=\kappa F_{\mathrm{gas}}\frac{M_{200}}{4\mathrm{\pi}g(C)r^2_{200}}\frac{g^{2}(Cr^\prime)}{r^{\prime\,2}}\:.
\end{equation}
The optical depth of the halo is then dependent on its properties, such as mass and concentration, and on the chosen dust opacities. We reiterate that in these simulations we assume and an absorption coefficient of $10^{5}\,\mathrm{cm}^{2}\,\mathrm{g}$, which for a Milky Way-\textit{like} dust-to-gas ratio $D=0.01$ leads to dust opacity of $\kappa=10^{3}\,\mathrm{cm}^{2}\,\mathrm{g}$ for both UV and IR radiation.

For an NFW halo with a given set of parameters there exists a critical luminosity above which the shell will be able to escape the halo. This escape condition is when the shell achieves a velocity larger than the peak of the circular velocity $v_{\mathrm{circ}}=\sqrt{g(Cr^\prime)/g(C)r^\prime}$. For the parameters of our fiducial NFW halo the circular velocity peaks at $r\approx2.163r_{\mathrm{s}}$ with a value $\approx0.047\,[C/g(C)]^2$. Setting $f[\tau]=1$ and left hand side of equation \ref{eq:eom2} to zero we obtain
\begin{equation}\label{eq:Lcrit}
\begin{split}
 L_{\mathrm{crit}}&=\frac{F_{\mathrm{gas}}v^{4}_{200}\,c}{G}\left[\frac{g(Cr^\prime)}{g(C)r^\prime}\right]^2 \\ &=0.047\left[\frac{C}{g(C)}\right]^2\frac{F_{\mathrm{gas}}v^{4}_{200}\,c}{G}\:.
\end{split}
\end{equation}
Substituting equation (\ref{eq:Lcrit}) into equation (\ref{eq:eom2}) we arrive at the dimensionless equation of motion for a thin shell expanding radially outward in spherically symmetric NFW halo
\begin{equation}\label{eq:eom3}
 \frac{d}{dt}\left[g(Cr^\prime)\frac{dr^\prime}{dt^\prime}\right]=0.047f[\tau]\frac{C^2}{g(C)}\xi-\frac{g^2(Cr^\prime)}{r^{\prime\,2}g(C)}\:,
\end{equation}
where $\xi=L/L_{\mathrm{crit}}$. We then numerically integrate equation (\ref{eq:eom3}) to produce the expected analytic result for the propagation of thin shell of gas radially outwards due to radiation pressure on dust evenly mixed within it. We note that we use the fiducial softening length of the gas particles as our initial $r^\prime$ choice when calculating the analytic result.

\section{Comparison to the analytic result}
\label{sec:sim_anlyt}
\begin{figure*}
 \includegraphics[width=1.025\textwidth]{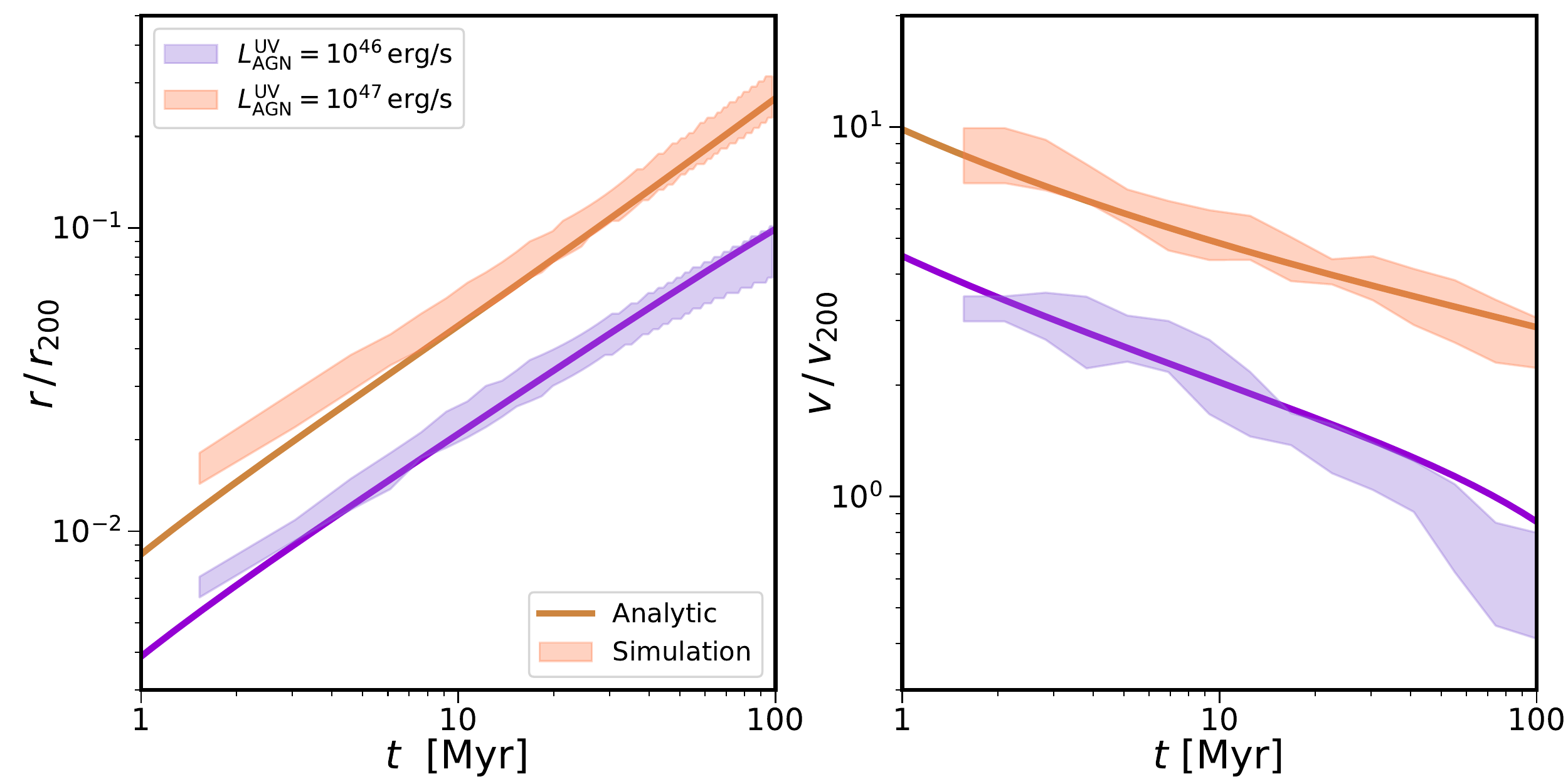}
 \caption{Shock position (left) and velocity (right) as a function of time for the simulated halo (filled region) and analytic model (solid line). The $10^{46}$ and $10^{47}\,\mathrm{erg}\,\mathrm{s}^{-1}$ bolometric luminosity AGN are denoted by the purple and orange regions, respectively. Both luminosities produce excellent agreement with the analytic solution. The propagation of the lower luminosity AGN shock begins to stall relative to the expected analytic result after $60\,\mathrm{Myr}$ due to finite numerical resolution and the confining pressure of the ambient medium of the halo gas, which is neglected in the analytic model.}
 \label{fig:UVonly}
\end{figure*}

\begin{figure}
 \includegraphics[width=\columnwidth]{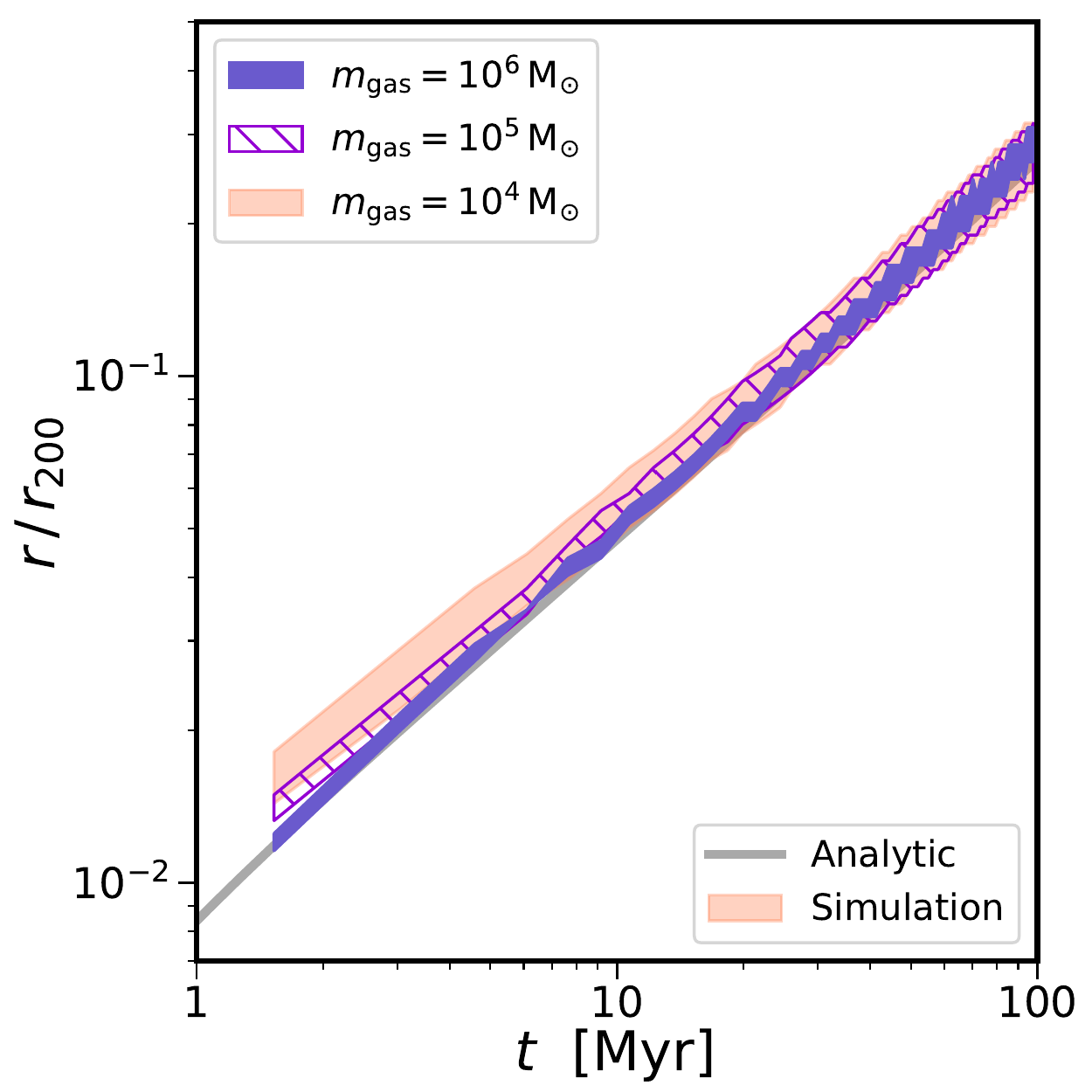}
 \caption{Shock position as a function of time for the simulated halo (filled region) and the analytic model (solid grey line). We compare the position of the shock front for three different numerical resolutions: $m_{\mathrm{gas}}=10^{4}$, $10^{5}$ and $10^{6}\,\mathrm{M}_{\astrosun}$, denoted by orange, purple hatch and blue regions, respectively. With increasing resolution the initial shock development occurs marginally faster. However, for $t\geq5\,\mathrm{Myr}$ there is excellent between the analytic result and all resolutions. As the resolution degrades the uncertainty in the shock position increases.}
 \label{fig:Con_mass}
\end{figure}

We now perform RHD simulations and compare to the expected analytic result. We mimic the analytic setup as closely as possible to ensure a fair comparison. We use a relaxed NFW halo initial conditions without a central galaxy. The black hole isotropically emits radiation at a constant luminosity for the $100\,\mathrm{Myr}$ duration of the simulation. Fig. \ref{fig:Outschm} shows the gas density, radial velocity, and UV and IR energy densities at $t=50\,\mathrm{Myr}$ for the NFW\_UVIR47 simulation. In this simplified setup the radiation drives a spherical outflow. The UV radiation declines radially from the source and the IR radiation becomes trapped behind the high-density material swept out by the outflow.

We begin with the single scattering regime, i.e. the injected UV radiation is absorbed by the dust but we neglect its re-emission in the IR band. We examine $2$ luminosities: $10^{46}$ and $10^{47}\,\mathrm{erg}\,\mathrm{s}^{-1}$, i.e. simulations NFW\_UV46 and NFW\_UV47. The critical luminosity required for the shell to escape the halo is $L_{\mathrm{crit}}=4.14\times10^{45}\,\mathrm{erg}\,\mathrm{s}^{-1}$. Therefore, both luminosities are super-critical and if the simulations were run for a sufficient period of time the shock would escape the halo.

In Fig. \ref{fig:UVonly} we compare the position and velocity of the simulated shock relative to $r_{200}$ and $v_{200}$, respectively,  as a function of time to the predicted analytic result for the two luminosities. In general, there is good agreement between the simulated and analytic results for the position of the shock front as a function of time. At early times the simulated shock is marginally ahead of the analytic model, but for $t>10\,\mathrm{Myr}$ there is good agreement between the position of the simulated shock and the position predicted by the analytic model. For the higher luminosity case, the velocity of the shock is in excellent agreement with the shock velocity predicted by the analytic result. 

The thin shell approximation appears to be more valid for the higher luminosity AGN. For the lower luminosity simulation the shock begins to stall and lags behind the predicted analytic result for $t>60\,\mathrm{Myr}$. This becomes more apparent when comparing the velocity of the simulated shock to the analytic result, where at late times the simulate shock is clearly moving more slowly than the analytic expectation. The discrepancy between the simulated and analytic results is due to two effects. First, the simulated shock front has a finite width compared to the thin shell approximation used in the analytic model, which is caused by the finite resolution of the simulation, the radiation's finite mean free path and thermal broadening of the shock. This reduces the density of the shock, reducing its optical depth and the momentum the radiation can impart on the gas. Second, the analytic solution neglects the confining pressure of the gas in the halo. The density contrast of the simulated shock falls with time and for $t>60\,\mathrm{Myr}$ the contrast drops below unity and the shell begins to stall due to pressure from the ambient medium of the halo.

\begin{figure*}
 \includegraphics[width=1.025\textwidth]{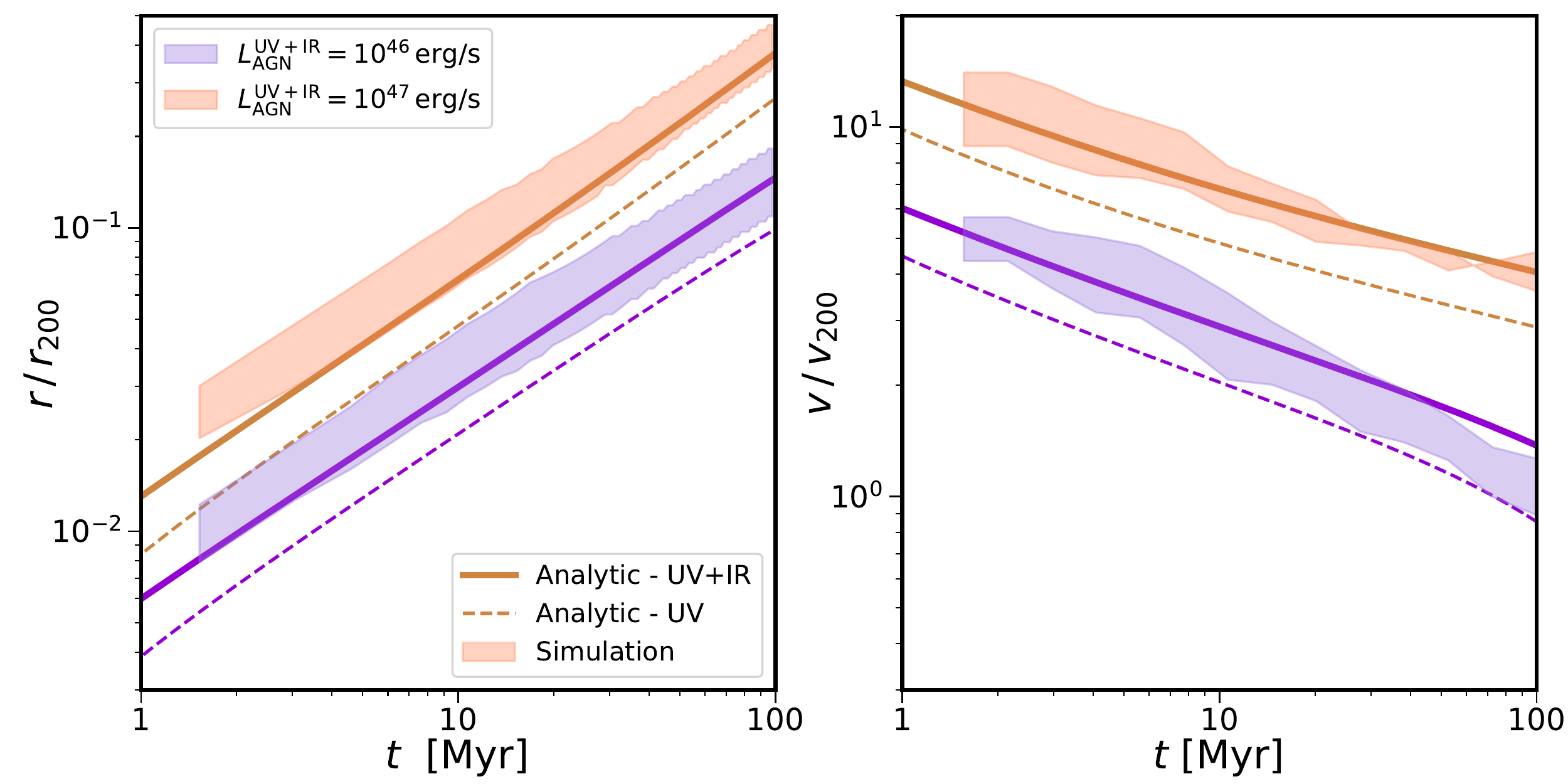}
 \caption{Shock position (left) and velocity (right) as a function of time for the simulated halo and the expected analytic result. Two AGN luminosities are shown: $10^{46}$ and $10^{47}\,\mathrm{erg}\,\mathrm{s}^{-1}$. Linestyles are the same as Fig. \ref{fig:UVonly}. In addition, we plot the UV only analytic result (dashed lines) to demonstrate the expected impact of the momentum boost associated with multiply scattered IR photons. We find excellent agreement between the simulations and the analytic model. The lower luminosity shock is impacted by the ambient medium of the halo.}
 \label{fig:UV+IR}
\end{figure*}

We now explore how the simulated shock propagation varies as a function of numerical resolution, using a central black hole isotropically emitting UV radiation with a luminosity of $10^{47}\,\mathrm{erg}\,\mathrm{s}^{-1}$ and neglecting IR reprocessing. We run three simulations: NFW\_UV47, NFW\_UV47\_m1e5 and NFW\_UV47\_m1e6, across which the numerical resolution changes by a factor of $100$. Each simulation setup is relaxed for $1\,\mathrm{Gyr}$ and have target gas masses of $m_{\mathrm{gas}}=10^{4}$, $10^{5}$ and $10^{6}\,\mathrm{M}_{\astrosun}$, respectively. These simulations have a corresponding softening lengths of $\epsilon=83$, $177$ and $365\,\mathrm{pc}$, respectively. We note that the number of neighbours into which the photons are injected remains constant between the simulations and so the radius of the injection sphere increases with increasing gas mass. However, we have examined the impact of changing the number of cells into which the photons are injected and found it has a negligible impact on the propagation of the shock front.

Fig. \ref{fig:Con_mass} demonstrates how the simulated shock position relative to $r_{200}$ as a function time evolves with numerical resolution. During the initial acceleration of the shock front $(t<5\,\mathrm{Myr})$ the shock front travels further with increasing numerical resolution, with the fiducial resolution slightly overestimating the shock position relative to the expected analytic result. However, after the initial acceleration all three resolutions are in agreement with each other and in excellent agreement with the analytic result. We note that as the resolution of the simulation decreases the noise in the position of the shock increases, making it harder to accurately determine the location of the shock front.

For a medium of sufficient optical depth, the IR photons can become trapped, scattering multiple times before they escape. We examine the additional momentum provided by multiply scattered photons by allowing absorbed UV photons to be re-emitted as IR photons. The trapping of IR photons leads to them diffusing out of the halo at an effective speed that is significantly lower than the speed of light. Using the characteristic trapping time-scale \citep[see][]{Costa2018a}, we can estimate the effective propagation speed via
\begin{equation}\label{eq:veff}
 v_{\mathrm{eff}}=\frac{c}{3\tau_{\mathrm{IR}}}\:.
\end{equation}
This diffusion speed implies that there is a maximum velocity with which the outflow can travel before IR photons no longer efficiently couple to the medium. For a typically observed outflow velocity of  $\sim1000\,\mathrm{km}\,/\mathrm{s}$, our chosen opacity the maximum optical depth of a halo before the coupling becomes inefficient is $\tau_{\mathrm{IR}}\approx100$.

The simulation setup consists of the relaxed adiabatic NFW halo with a central black hole isotropically emitting UV radiation at constant bolometric luminosities of $L_{\mathrm{AGN}}=10^{46}$ and $10^{47}\,\mathrm{erg}\,\mathrm{s}^{-1}$, i.e. simulations NFW\_UVIR46 and NFW\_UVIR47. UV photons that are absorbed by the dust are re-emitted in the IR band. Fig. \ref{fig:UV+IR} compares the simulated shock position and velocity relative to $r_{200}$ and $v_{200}$, respectively, as a function of time for simulations with reprocessing and IR multi-scattering. We show the expected analytic result with and without IR multi-scattering to demonstrate the additional momentum coupled to the gas by the IR photons. The simulated shock position shows excellent agreement with the analytic result that includes the multi-scattering of IR photons. The velocity of the simulated shock is also in good agreement with the expected result, though somewhat noisy. There are some hints that the simulated lower luminosity shock is decelerating faster than expected, which is due to the finite numerical resolution, thermal broadening of the shock and the confining pressure of the ambient gas of the halo.

\begin{figure}
 \includegraphics[width=\columnwidth]{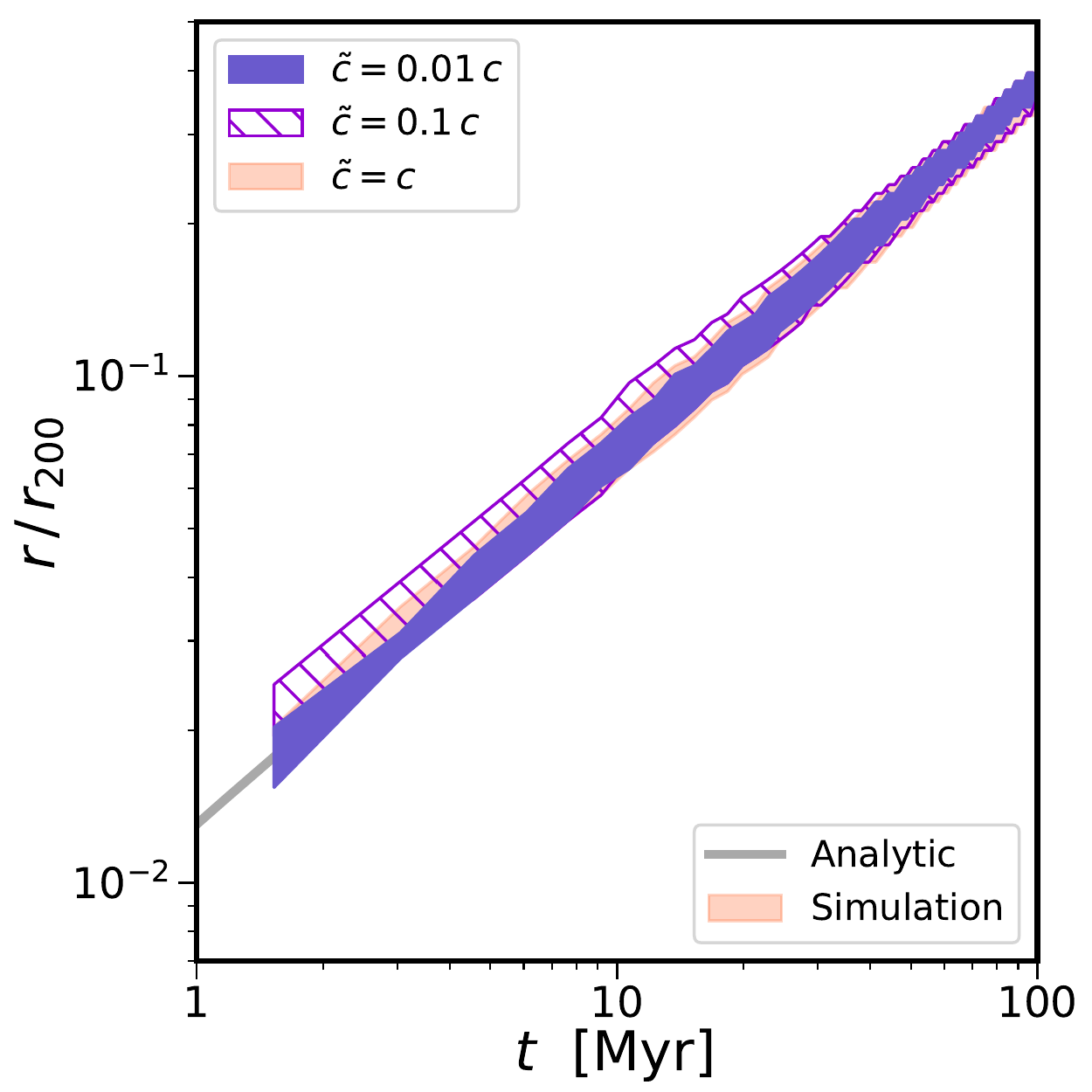}
 \caption{Shock position as function of time for the simulated halo (filled region) and the analytic model (solid grey line). We compare the position of the shock front for three reduced speed of light approximations: $\tilde{c}=0.01c$, $0.1c$ and $c$, denoted by the blue, purple hatch and orange regions, respectively. Irrespective of the chosen speed of light approximation we find that the simulated shock position is in good agreement with the analytic expectation.}
 \label{fig:Con_clight}
\end{figure}

As shown in equation \ref{eq:veff}, the efficient coupling of momentum from IR photons requires that the propagation velocity of the shock be lower than the ratio of the speed of light over the optical depth of the halo. Therefore, use of the reduced speed of light approximation in the simulations is potentially a concern, too low a speed of light and the simulation will not capture the additional momentum boost from the IR photons. In this work, our fiducial speed of light is $(\bar{c}=0.1c)$ in an NFW halo with an initial optical depth $\tau=11.37$, implying a maximum outflow velocity of $v_{\mathrm{out}}=8.8\times10^{3}\,\mathrm{km}\,/\mathrm{s}$. To explore the impact of the reduced speed of light approximation we run $3$ simulations of a central black hole isotropically emitting UV radiation at a bolometric luminosity of $10^{47}\,\mathrm{erg}\,\mathrm{s}^{-1}$, allowing absorbed UV radiation to be reprocessed into IR photons, i.e. NFW\_UVIR47\_c0p01, NFW\_UVIR47 and NFW\_UVIR47\_c1p00. The reduced speed of light approximation is set to $\bar{c}=10^{-2}c$, $10^{-1}c$ and $c$, respectively. This implies maximum outflow velocities of $v_{\mathrm{out}}=8.8\times10^{2}$, $8.8\times10^{3}$ and $8.8\times10^{4}\,\mathrm{km}\,/\mathrm{s}$, respectively, before the coupling of IR photons becomes inefficient.

Fig. \ref{fig:Con_clight} shows the impact of varying the reduced speed of light approximation on the position of the shock front relative to $r_{200}$ as a function of time for the simulations and the analytic solution. The simulated results are well converged and in good agreement with the analytic result, with a small amount of noise in the early time $(t\leq5\,\mathrm{Myr})$ simulated shock position. Therefore, we conclude that the results presented in this work are robust to our chosen reduced speed of light approximation of $\tilde{c}=0.1c$. We note that for the $\tilde{c}=10^{-3}c$ the simulation very rapidly fails because the photons travel at the same speed as the outflowing mesh generation points, which results in large advection errors.

\section{Outflow properties}
\label{sec:outprops}
\begin{figure*}
 \includegraphics[width=\textwidth]{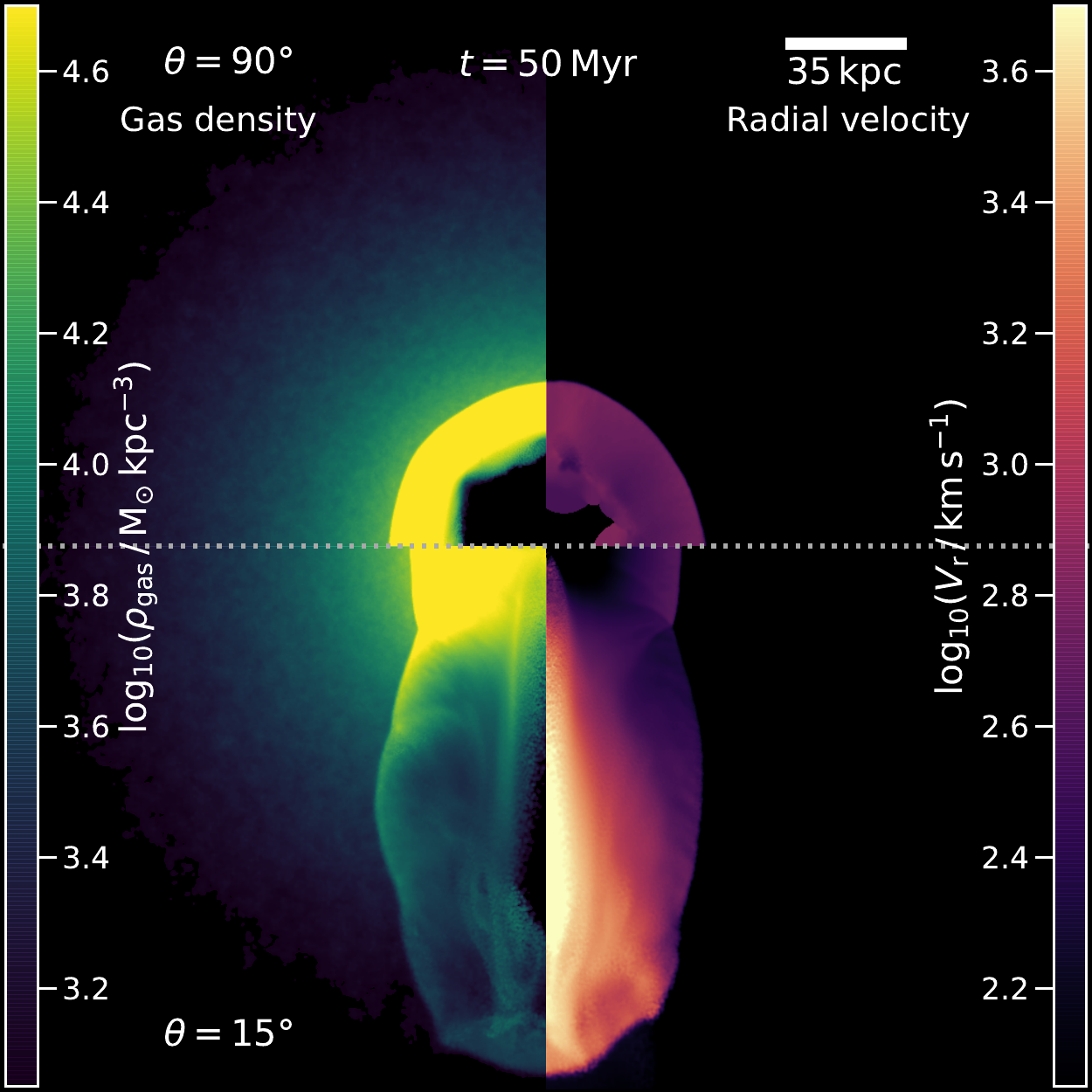}
 \caption{Black hole centred gas density (left) and radial velocity (right) slices of $20\,\mathrm{kpc}$ depth through two RHD simulations at $t=50\,\mathrm{Myr}$. Radiation injection opening angles of $90\degree$ (top) and $15\degree$ (bottom) are shown. A smaller opening angle produces more collimated outflows with larger radial velocities, as the radiation couples its momentum to a smaller amount of gas.}
 \label{fig:BiPschm}
\end{figure*}

We now focus on the properties of the outflow generated by the radiation injection. Recent work by \citet{Ishibashi2018} has shown that the radiatively driven outflows in a $1$D isothermal potential have properties that are reasonably well matched to the average properties of observed outflows. However, the model did not reproduce the extreme values observed in some outflows. To make the problem tractable the analytic models make several simplifying assumptions. They assume that the radiation injection and the resulting outflow are spherically symmetric, i.e. $1$D in nature, and that the outflow sweeps all of the available material into a thin shell as it propagates outwards. These assumptions impact the properties of the outflows predicted. Previous work has shown that increasing the dimensionality of the problem by one, i.e. $2$D, and accounting for a central galaxy can lead to very different outflow properties \citep{Hartwig2018}. The initial shock creates a chimney in the direction of the lower density medium perpendicular to the galaxy through which the injected energy preferentially escapes. The outflow contains less material, but it is moving at a greater velocity.

\begin{figure}
 \includegraphics[width=\columnwidth]{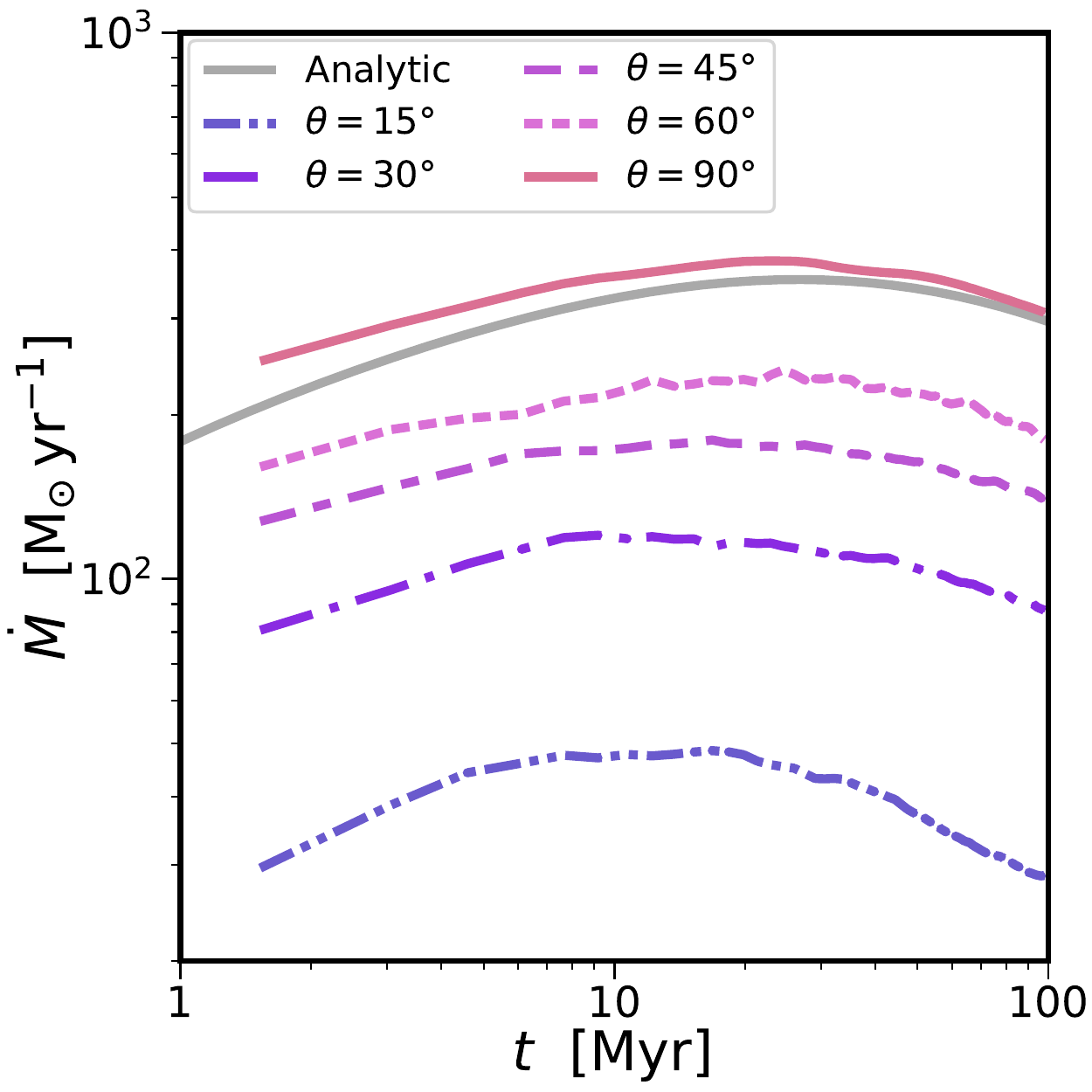}
 \caption{Mass outflow rate as a function of time for a black hole with a bolometric luminosity $L_{\mathrm{AGN}}=10^{47}\,\mathrm{erg}\,\mathrm{s}^{-1}$. We compare the analytic result (grey solid line) to simulated results with an opening angle of $\theta=15\degree$ (blue dash-dot-dot), $30\degree$ (purple dash-dot), $45\degree$ (light purple dash-dash), $60\degree$ (pink dashed) and $90\degree$ (red solid). For isotropic injection ($90\degree$) the simulated result mildly over-predicts the mass outflow rate relative to the expected analytic result, but the result converges at lates. As the injection opening angle is reduced and the radiation becomes more collimated the mass outflow reduces, as expected, and the peak outflow rate shifts to earlier times.}
 \label{fig:Mdot}
\end{figure}

Observationally, an outflow's properties are typically characterized by three quantities: its mass outflow rate, $\dot{M}_{\mathrm{out}}$, its momentum flux, $\dot{P}_{\mathrm{out}}$ and its kinetic power, $\dot{E}_{\mathrm{kin}}$ \citep[e.g.][]{Maiolino2012,Gonzalez-Alfonso2017,Fiore2017}. Each observation provides a snapshot of the outflow, measuring it properties at a single point in time. Following previous work, we define these key outflow properties as
\begin{equation}\label{eq:Mout}
 \dot{M}_{\mathrm{out}}=\frac{Mv}{r}\:,
\end{equation}
\begin{equation}\label{eq:Pout}
 \dot{P}_{\mathrm{out}}=\frac{Mv^{2}}{r}\:,
\end{equation}
\begin{equation}\label{eq:Ekin}
 \dot{E}_{\mathrm{kin}}=\frac{1}{2}\frac{Mv^{3}}{r}\:,
\end{equation}
where $M$ is the mass of material in the outflow, $v$ is the radial velocity of the outflowing material and $r$ is the halocentric radial distance of the outflow material. As noted in Section \ref{sec:outflowdef}, we define material in the simulations as outflowing if its halocentric radial velocity is greater than $50\,\mathrm{km}\,\mathrm{s}^{-1}$.

All numerical simulations in this Section use the same setup, an NFW halo with an external static dark matter potential and an adiabatic gas halo. At the centre of the halo is a black hole emitting UV radiation with a constant luminosity of $L_{\mathrm{AGN}}=10^{47}\,\mathrm{erg}\,\mathrm{s}^{-1}$ throughout the $100\,\mathrm{Myr}$ of the simulation. Photons are injected in a biconical region with an opening angle $\theta$ that is oriented parallel to the $z$-axis. We examine a range of values for the opening angle: $\theta=15\degree$, $30\degree$, $45\degree$, $60\degree$ and $90\degree$, where $90\degree$ is equivalent to spherical injection. These simulations are labelled NFW\_UV47\_$\theta$15, NFW\_UV47\_$\theta$30, NFW\_UV47\_$\theta$45, NFW\_UV47\_$\theta$60 and NFW\_UV47, respectively. The bipolar cone always contains $256$ gas cell neighbours and each cell receives a fraction of the emitted photons based on the fraction of the solid angle it covers relative to the total solid angle covered by the selected neighbours. We confirm that the number of neighbours the photons are injected into has a negligible impact on the results presented.

If IR radiation is neglected, then the momentum flux of any outflow driven by UV radiation pressure is limited to $L_{\mathrm{AGN}}\,/\,c$. Hence, the ratio of the momentum of the outflow to the AGN input is limited to $\zeta=c\dot{P}_{\mathrm{out}}\,/L_{\mathrm{AGN}}\leq1$. Typical momentum ratios of observed outflows are in the range $\zeta\approx1-30$ \citep{Cicone2014,Zakamska2014,Carniani2015,Zakamska2016,Fiore2017}. Therefore, if radiation pressure is the mechanism driving the outflow the simulations will not reach the observed values. However, radiation produces a shock and the hot material of the outflow will adiabatically expand as it propagates through the halo. The expansion of the outflow material will do `$PdV$' work on its surroundings. For a sufficiently strong shock the contribution of the expanding gas to the momentum flux and kinetic power of the outflow becomes non-negligible, potentially bringing the simulated momentum ratio towards observed values. In this Section we explore the contribution of the expanding outflow by collimating the radiation. The radiation then couples its momentum to a smaller quantity of gas, which drives an outflow with greater velocity and produces a stronger shock. To simplify this study we neglect IR radiation, and its associated momentum boost, to examine the impact of collimation. We note that, due to the chosen IR opacity, any momentum boost due to IR radiation would be significantly larger than expected in observed systems, which would limit the reliability of any comparison which includes IR radiation.

\begin{figure}
 \includegraphics[width=\columnwidth]{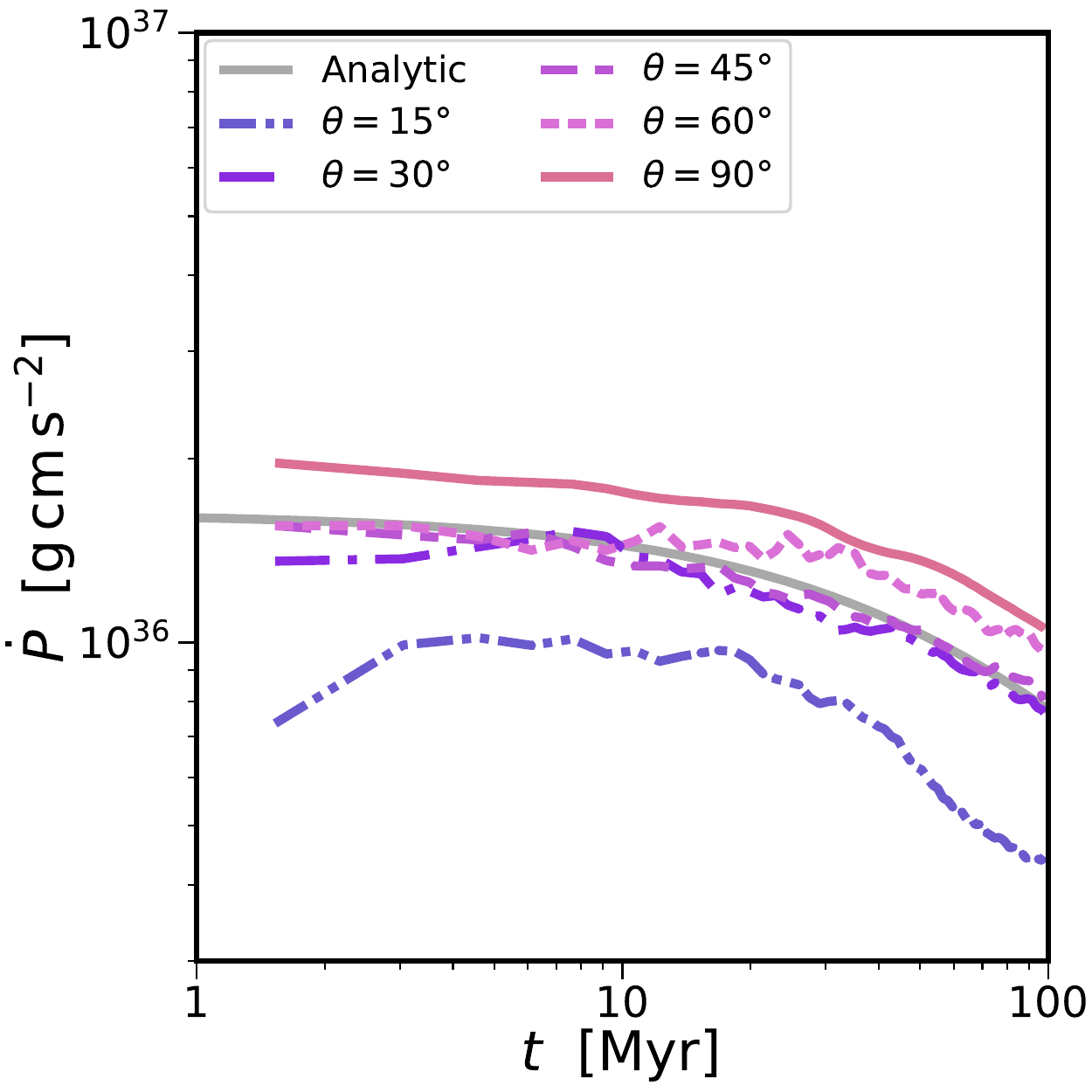}
 \caption{Momentum flux of the outflow as a function of time for a black hole with a bolometric luminosity $L_{\mathrm{AGN}}=10^{47}\,\mathrm{erg}\,\mathrm{s}^{-1}$. Line styles are the same as Fig. \ref{fig:Mdot}. The simulated momentum flux evolves with time in a manner consistent with the expected analytic result, a steady decrease with time, but the normalization is a  consistently $15$ per cent higher than the analytic model. This is driven by the finite width of the shock and the associated distribution in velocities. As the radiation is collimated the momentum flux decreases, but the increased outflow velocity compensates, to an extent, for its reduced mass.}
 \label{fig:Pdot}
\end{figure}

In Fig. \ref{fig:Mdot} we plot the mass outflow rate as a function of time for the simulations with various opening angles and the expected analytic result. The simulation with isotropic radiation injection, $\theta=90\degree$, has a mass outflow rate that is in reasonable agreement with the analytic solution. The evolution of the mass outflow rate with time has a very similar shape, peaking at $t\sim30\,\mathrm{Myr}$. The normalization of the simulated mass outflow rate is marginally higher than the analytic result, but the values converge at late times. At early times $(t=1.5\,\mathrm{Myr})$, the simulated mass outflow is $12$ per cent larger than the analytic solution. As shown in the previous Section, the shock initially develops more rapidly than the analytic result and its outflow will therefore initially contain more mass. At late times, the positions of the shocks are in good agreement, but the simulated shock has a finite width compared to the thin shell approximation of the analytic solution. This width leads to the simulated shock sweeping up material that has not been encountered by the analytic shock, resulting in a mass outflow rate that is marginally higher than expected.

As expected, the mass outflow rate decreases as the opening angle of the injection region is reduced. Relative to the isotropic injection case, the mass outflow rate at $t=100\,\mathrm{Myr}$ is $42$, $55$, $71$ and $91$ per cent lower for $\theta=60\degree$, $45\degree$, $30\degree$ and $15\degree$, respectively. The collimation of the radiation naturally reduces the amount of gas it can impact, but the reduction in the mass outflow is smaller than the naive expectation of  $50$, $70$, $87$ and $97$ per cent from the ratio of the solid angle subtended by the radiation relative to the isotropic injection case. This is partly driven by the collimated outflows generating radial inward motions in the plane perpendicular to the outflow direction, which results in more material being drawn into the outflow than expected. There is evidence that the peak mass outflow rates shifts from $t\sim30\,\mathrm{Myr}$ for isotropic injection to $t\sim10\,\mathrm{Myr}$ for an injection region with a $15\degree$ opening angle. This is driven by the competition between the mass entrained in the outflow, its velocity and its radial extent, for a smaller opening angle the mass loading is reduced but the velocity and radial extent of the outflow are larger at fixed time.

\begin{figure}
 \includegraphics[width=\columnwidth]{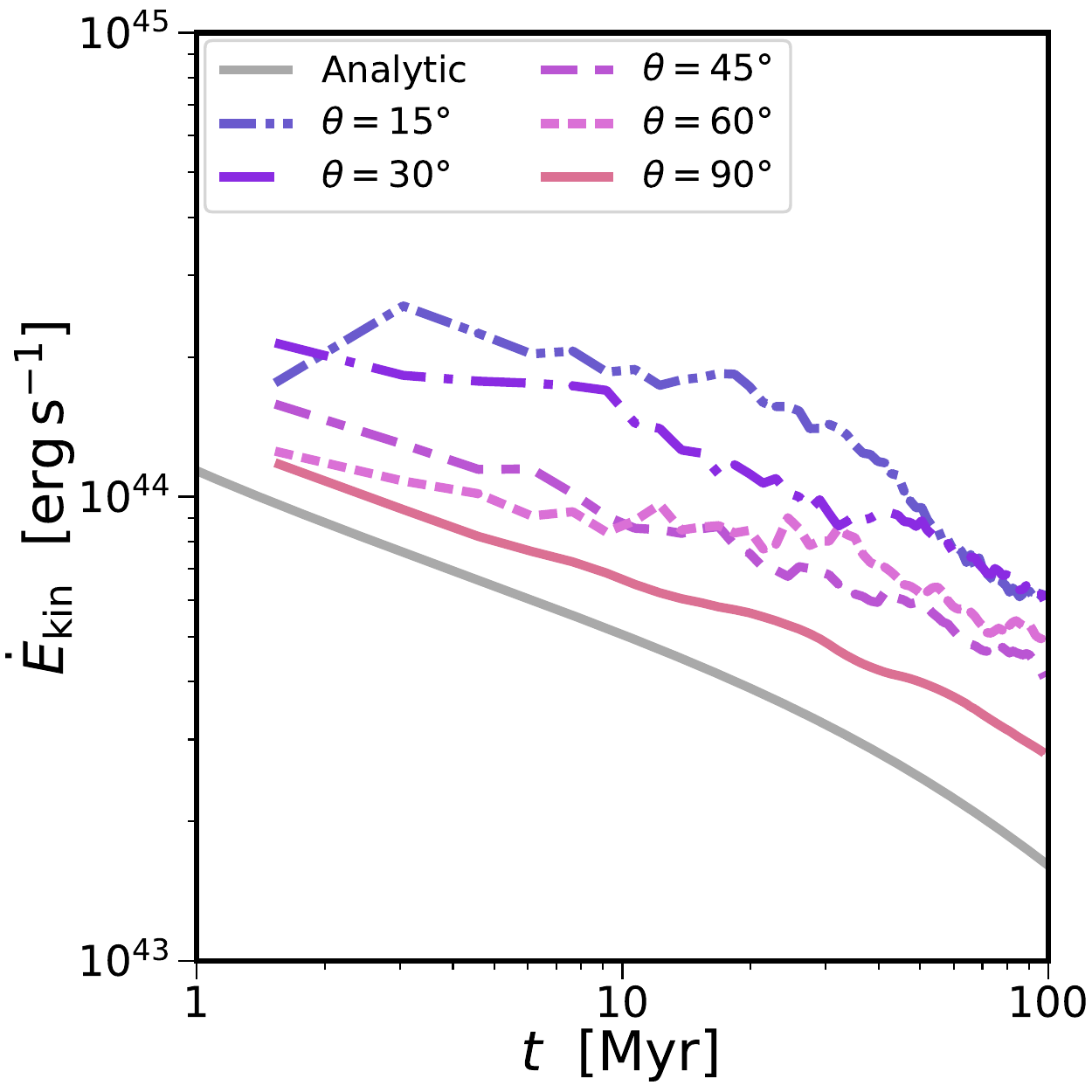}
 \caption{Kinetic power of the outflow as a function of time for a black hole with a bolometric luminosity $L_{\mathrm{AGN}}=10^{47}\,\mathrm{erg}\,\mathrm{s}^{-1}$. Line styles are the same as Fig. \ref{fig:Mdot}. For isotropic injection, the kinetic power of the simulated outflow is larger than the analytic expectation due to the velocity distribution of the outflow, driven by its finite width relative to the thin shell approximation of the analytic model. As the opening angle of radiation injection decrease the kinetic power increases due to its increased velocity.}
 \label{fig:Ekin}
\end{figure}

We plot the momentum flux as a function of time for the simulated and analytic outflows in Fig. \ref{fig:Pdot}. For isotropic injection, the simulated and analytic results have very similar evolutions, with the momentum flux of the outflow slowly declining with time. However, the simulated result is consistently $15$ per cent higher than the analytic expectation. This is driven by two effects. First, the mass outflow rate of the simulation is slightly higher than the analytic result, especially at early times. Second, the finite width of the simulated shock front leads to a range of velocities and this results in a higher momentum flux at late times. As the opening angle of the injection region reduces we see a reduction in the momentum flux relative to isotropic, but the values are very similar for $\theta=60\degree$, $45\degree$ and $30\degree$. The momentum flux is a product of the mass outflow rate and the outflow velocity. As the radiation becomes more collimated the momentum is injected into a small amount of gas. However, the radiation accelerates this gas to larger velocities. For these three opening angles the two competing effects offset each other. For the smallest opening angle, $\theta=15\degree$, the momentum flux is a factor of $2$ lower than the isotropic case because the mass outflow rate is smaller than the boost in velocity. 

The largest momentum ratio is achieved by the isotropic injection case, with $\zeta=0.59$. As the opening angle decreases the momentum ratio decreases, reducing by $48$ per cent to $\zeta=0.31$ for an opening angle of $15\degree$. The simulated outflows remain momentum-driven and the adiabatic expansion of the outflow provides a negligible contribution. Therefore, we conclude that collimating the radiation does not lead to a significant change in the strength of the shock generated by the radiative feedback in this idealized NFW halo setup. Haloes that lack sufficient optical depth for IR photons to become trapped and scatter multiple times will not achieve the momentum ratios of $\zeta=1-30$ typically seen in observed outflows \citep{Cicone2014,Zakamska2014,Carniani2015,Fiore2017}.

The kinetic power of the simulated and analytic outflows as a function of time are shown in Fig. \ref{fig:Ekin}. The analytic result and isotropic injection simulation outflows have a similar kinetic power evolution with time, decreasing by a factor of $5$. The simulated result is initially $17$ per cent higher at $t=1.5\,\mathrm{Myr}$ and this difference increases mildly with time. This is caused by the numerical resolution of the simulation and thermal broadening of the shock that results in it having a finite width, compared to the thin shell approximation of the analytic result. This leads to material in the shock having a range of velocities and the outflow having a larger kinetic power relative to the analytic result.

\begin{figure}
 \includegraphics[width=\columnwidth]{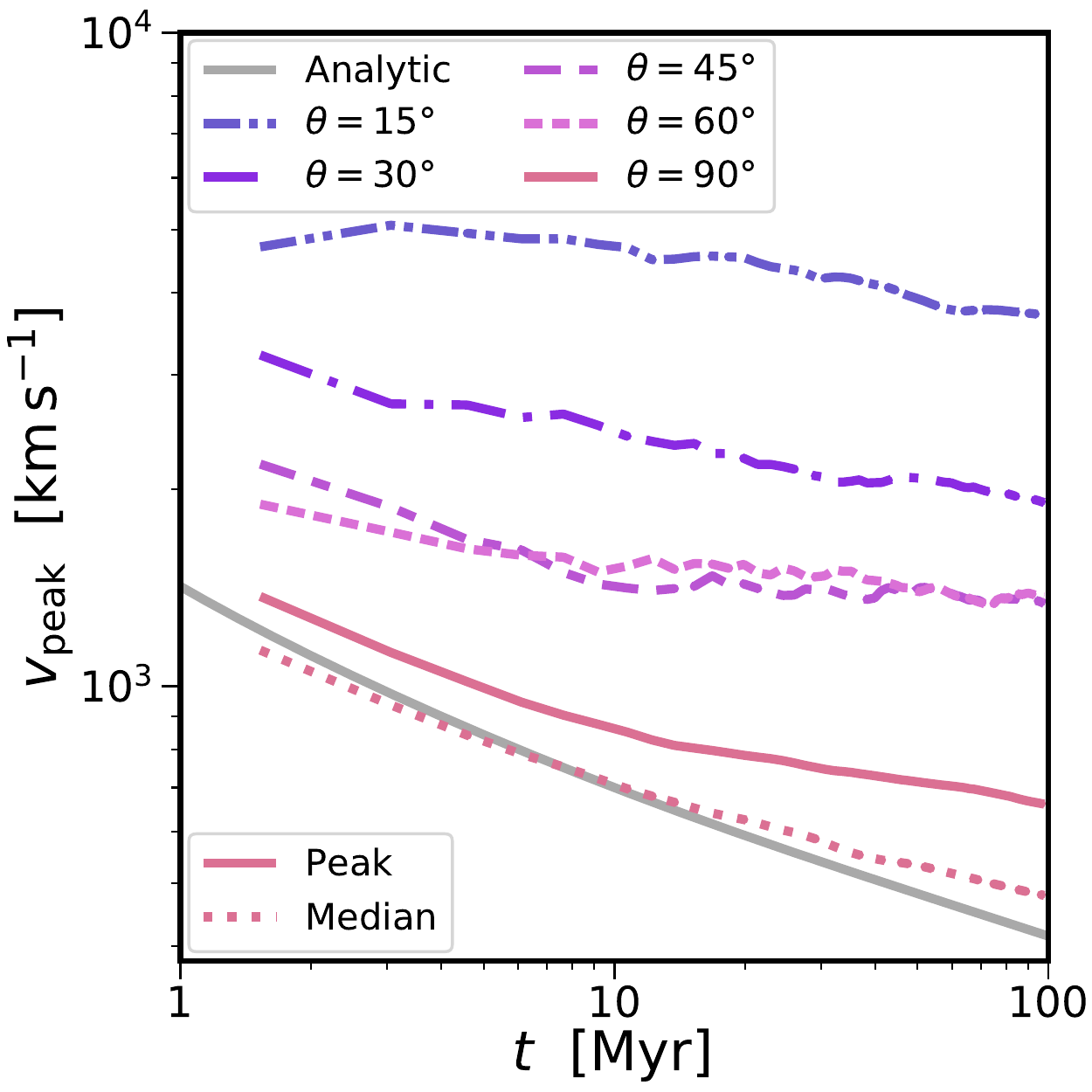}
 \caption{Peak velocity of the outflow as a function of time for a black hole with a bolometric luminosity $L_{\mathrm{AGN}}=10^{47}\,\mathrm{erg}\,\mathrm{s}^{-1}$. Line styles are the same as Fig. \ref{fig:Mdot}. In the isotropic injection case $(\theta=90\degree)$ the peak velocity of the simulated outflow is larger than the analytic expectation, due to the finite width of the outflow. However, if we consider the median velocity of the outflow we find good agreement with the analytic result. As expected, as the injection opening angle decreases the peak velocity increases because the radiation couples a greater amount of momentum per unit gas mass.}
 \label{fig:Vpeak}
\end{figure} 

For decreasing opening angle of the radiation injection region we find that the kinetic power of the outflow increases at fixed time. As the radiation is collimated it injects momentum into a smaller amount of gas, leading to significantly higher outflow velocities. As the opening angle is reduced the kinetic power of the outflow increases due to its $v^{3}$ dependence. The simulations achieve a peak energy ratio $\epsilon_{\mathrm{k}}=\dot{E}_{\mathrm{kin}}\,/L_{\mathrm{AGN}}=0.25$ per cent for the an opening angle $\theta=15\degree$ at $t=3\,\mathrm{Myr}$. This compares to typical observed values of $\epsilon_{\mathrm{k}}=0.3-3$ per cent \citep{Cicone2014,Zakamska2014,Carniani2015,Zakamska2016,Fiore2017}. Although the simulations may be capable of reproducing the lower end of the observed range of values, they cannot account for the high values in this idealized NFW setup.

In Fig. \ref{fig:Vpeak} we compare the peak outflow velocity as a function of time. For the analytic solution the outflow consists of a thin shell of material all propagating radially outwards at the same velocity. For the simulations we calculate the peak outflow velocity, $v_{\mathrm{peak}}$, as the $95^{\mathrm{th}}$ percentile of the velocity distribution for all material classified as outflowing, i.e. gas with a halocentric radial velocity component greater than $50\,\mathrm{km}\,\mathrm{s}^{-1}$. For isotropic radiation injection the finite width of the simulated shock leads to a distribution of the velocities, and a peak velocity that is larger than predicted by the analytic solution. As the width of the shock increases at late times the analytic and simulated peak velocity values diverge. If we instead examine the median velocity of the outflow material we find that the isotropic injection simulation yields an excellent agreement with the expected value from the analytic solution. As the opening angle of the radiation injection region decreases the peak velocity of the outflow increases, as the radiation is coupling momentum to a smaller amount of gas and producing larger acceleration gradients. The smallest opening angle, $\theta=15\degree$, achieves a maximum peak velocity of $5.1\times10^{3}\,\mathrm{km}\,\mathrm{s}^{-1}$. The simulated peak velocities are in reasonable agreement with the typical peak velocities of observed outflows \citep{Fiore2017}.

In general, we find reasonable agreement between the simulated outflow properties and those predicted by the analytic model. However, the finite width of the shock front leads to small discrepancies. For the simplified setup of an isolated NFW halo with no central galaxy, adiabatic gas and a constant dust-to-gas ratio of $1$ per cent, the simulated and analytic results yield outflows whose momentum flux and kinetic power are typically smaller than observed. If we collimate the radiation we find that kinetic power of radiatively driven feedback can explain the low-end range of the kinetic power of observed outflows, but the mass outflow rate and momentum flux of the outflows are further still from typically observed values. However, the simplifying assumptions are very strong and in the next Section we explore the impact of a central galaxy and a self-consistent dust distribution on the outflow properties. 

\section{Dust Physics}
\label{sec:dustphys}
\begin{figure}
 \includegraphics[width=\columnwidth]{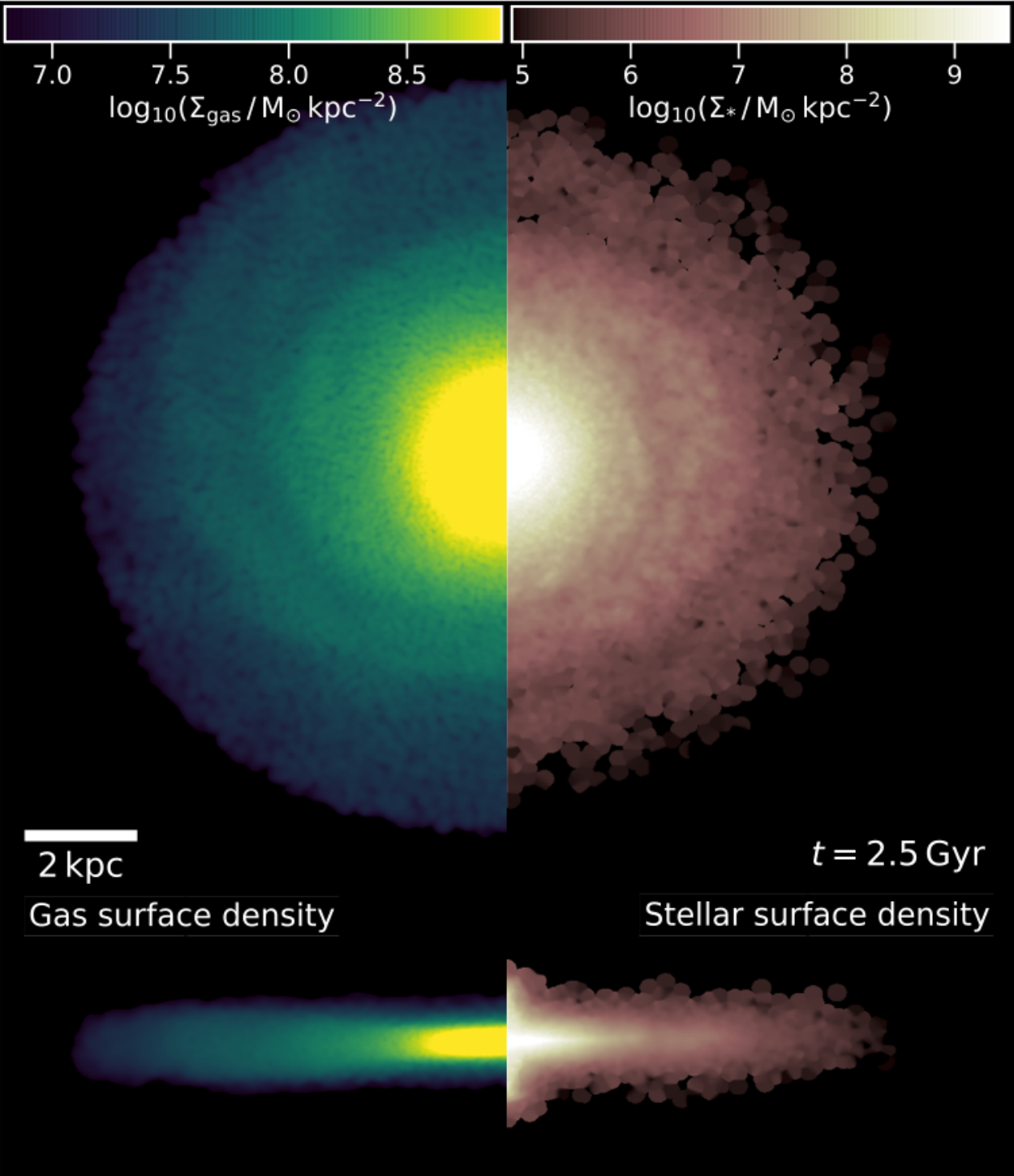}
 \caption{Face-on (top) and edge-on (bottom) gas (left) and stellar (right) surface densities of the NFW halo after the galaxy formation processes have been active for $2.5\,\mathrm{Gyr}$. The projection has a depth of $20\,\mathrm{kpc}$ and is centred on the black hole. This disc galaxy provides the initial conditions for simulations with a disc galaxy in an NFW halo.}
 \label{fig:Gal_sch}
\end{figure}

All results presented in the previous Sections assumed that dust grains are uniformly mixed within the halo gas with a constant dust-to-gas ratio $D=0.01$. Previous analytic and numerical work have all made this assumption \citep[e.g.][]{Murray2005,Thompson2015,Ishibashi2015,Ishibashi2016,Ishibashi2018,Costa2018a,Costa2018b,Hartwig2018}. In addition, the vast majority of work assumes the halo gas follows an NFW profile. Although these assumptions make the problem analytically tractable, they are strong assumptions that dramatically affect the optical depth of the halo and, therefore, the properties of radiatively driven outflows. To explore the impact of these assumptions we now model radiatively driven feedback in a halo that has cooled to form a central disc galaxy. During the formation of the galaxy, we model the formation of dust grains following \citetalias{McKinnon2017}. We note that all simulations throughout this Section have a target gas mass of $10^{5}\,\mathrm{M}_{\astrosun}$.

\begin{figure*}
 \includegraphics[width=\textwidth,keepaspectratio=True]{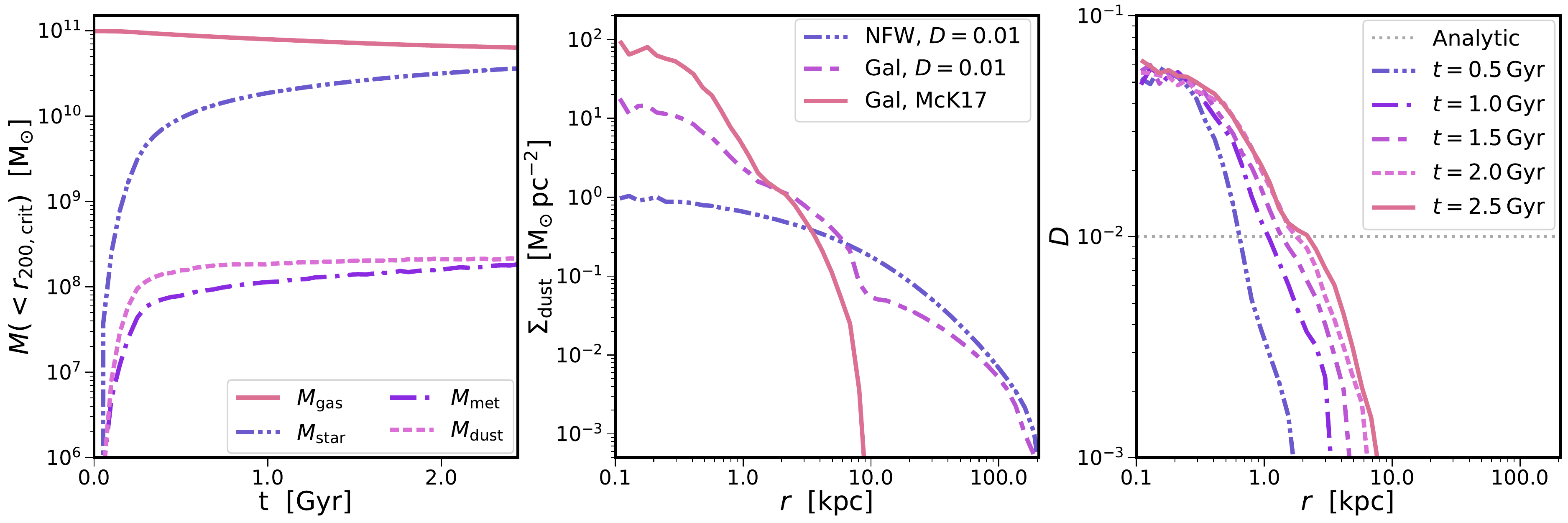}
 \caption{Properties of the initial disc galaxy. \textit{Left}: Gas (red solid), stellar (blue dash-dot-dot), metal (purple dash-dot) and dust (pink dashed) mass components with $r_{200,\mathrm{crit}}$. The global halo properties are well matched to a Milky Way-\textit{like} galaxy. \textit{Centre}: Radial dust mass surface density profile for the relaxed NFW halo with $D=0.01$ (blue dash-dot-dot), the NFW halo with a disc galaxy and $D=0.01$ (light purple dash-dash) and the NFW halo with a disc galaxy and self-consistently modelled dust (red solid). \textit{Right}: Dust-to-gas ratio radial profile as a function of time for the NFW halo with a central disc galaxy and self-consistently modelled dust. We plot $t=0.5$ (blue dash-dot-dot), $1.0$ (purple dash-dot), $1.5$ (light purple dash-dash), $2.0$ (pink dashed) and $2.5\,\mathrm{Gyr}$ (red solid). Dust is confined to the disc due to the lack of feedback and satellite mergers during its formation.}
 \label{fig:ICprops}
\end{figure*}

\subsection{Initial galaxy properties} 
As outlined in Section \ref{sec:meth} we evolved the relaxed, initial NFW halo with a live dark matter halo for a further $2.5\,\mathrm{Gyr}$ without injecting radiation. During this time the physical processes associated with galaxy formation are active, enabling the gas to cool radiatively and form stars. As the stars evolve they return mass and metals to their surrounding interstellar medium. Following \citetalias{McKinnon2017} , a fraction of the returned metals are in the form of dust drains. These grains are able to further grown via collisions with gas phase metals. Dust grains are destroyed by shocks and thermal sputtering. As shown in Fig. \ref{fig:Gal_sch}, the end result of this run is an NFW halo with a central Milky Way-\textit{like} disc galaxy.

In the left panel of Fig. \ref{fig:ICprops} we plot the mass of the baryonic components, i.e. gas, stars, metals and dust, of the halo with $r_{200}$ as a function of time. After $2.5\,\mathrm{Gyr}$ a stellar disc with a mass $3.63\times10^{10}\,\mathrm{M}_{\astrosun}$ has formed in the $xy$ plane of the halo with a radial extent of $8\,\mathrm{kpc}$. Over the first $0.5\,\mathrm{Gyr}$ a dust mass $M_{\mathrm{dust}}=1.4\times10^{8}\,\mathrm{M}_{\astrosun}$ rapidly forms, after which the dust mass grows much more slowly and the final halo has a dust mass $M_{\mathrm{dust}}=1.1\times10^{8}\,\mathrm{M}_{\astrosun}$. The build-up of gaseous metals initially occurs at a slower rate, but after $2.5\,\mathrm{Gyr}$ the dust and metal content of the halo is roughly similar, with a cumulative dust-to-metal ratio of $1.15$ within $r_{200}$.

\begin{figure}
  \includegraphics[width=\columnwidth]{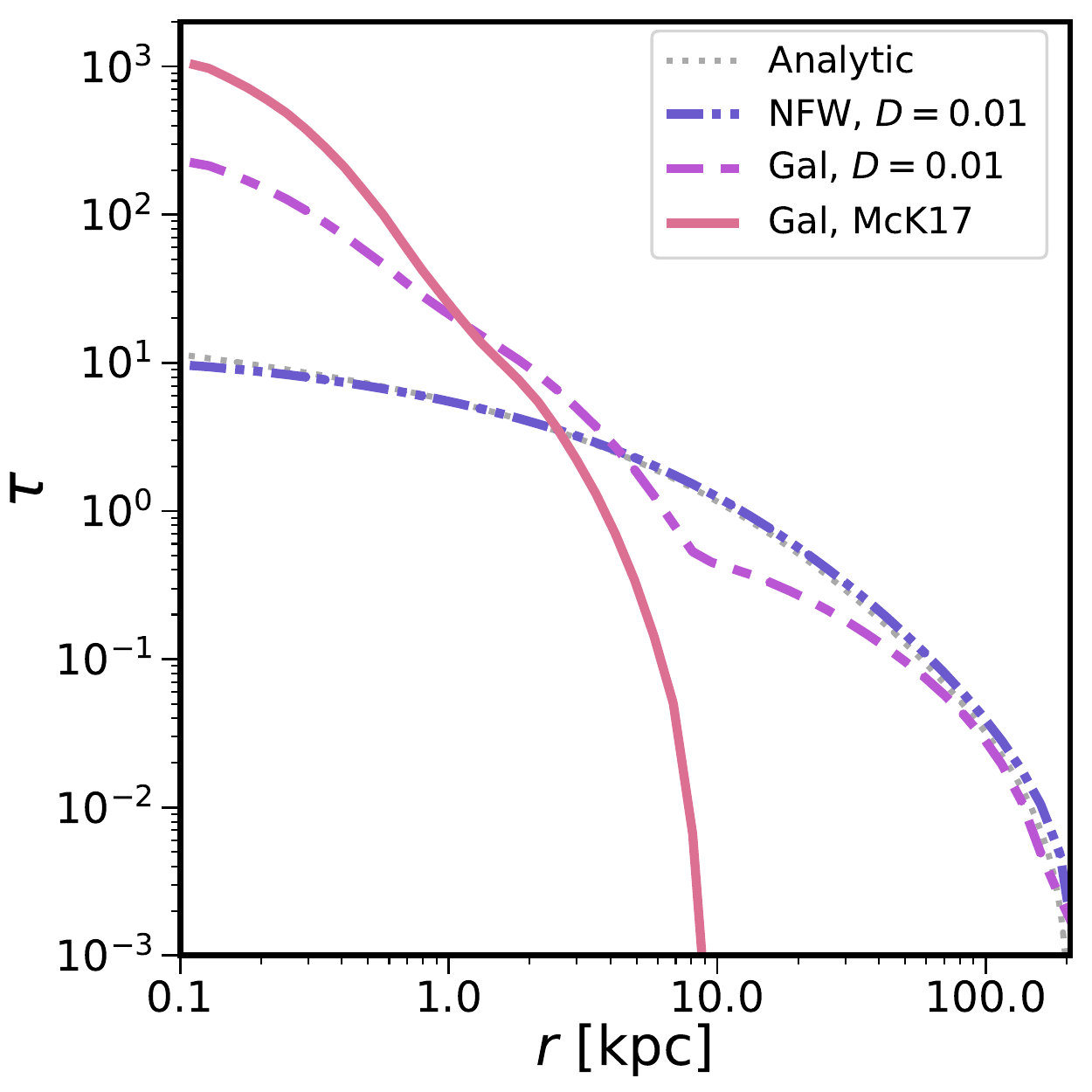}
 \caption{Optical depth integrated to $r_{200,\mathrm{crit}}$ as a function of radius for the NFW halo with a constant dust-to-gas ratio $D=0.01$ (blue dash-dot-dot), and NFW halo with a disc galaxy and constant dust-to-gas ratio $D=0.01$ (light purple dash-dash) and an NFW halo and a central disc galaxy and self-consistently modelled dust distribution (red solid). The formation of the galaxy leads to a factor $20$ increase in the central optical depth of the halo, assuming a dust-to-gas ratio $D=0.01$. Modelling the formation of dust leads to further increase in the optical depth by a factor of $4.5$, and yielding a final value of $\tau=1.04\times10^{3}$.}
 \label{fig:gal_tau} 
\end{figure}

The formation of the disc galaxy leads to a contraction of the halo. The central panel of Fig. \ref{fig:ICprops} shows the radial dust surface density profile, where the halo has been projected along the $z$ axis of the simulation (i.e. the galaxy is face on). For the initial NFW halo with a constant dust-to-gas ratio $D=0.01$ the surface density of the dust decreases strongly with radius from $1.34\,\mathrm{M}_{\astrosun}\,\mathrm{pc}^{-2}$ at $r=0.1\,\mathrm{kpc}$ to $5\times10^{-4}$ at $r_{200}$. After the formation of the disc, still assuming $D=0.01$, central dust surface density of dust at $0.1\,\mathrm{kpc}$ increases to $17.7\,\mathrm{M}_{\astrosun}\,\mathrm{pc}^{-2}$. However, due to material condensing to form the disc galaxy the dust surface density is lower than the initial NFW halo beyond $8\,\mathrm{kpc}$. Modelling the formation of dust we find that its central surface density increases still further, reaching a dust surface density of $95.9\,\mathrm{M}_{\astrosun}\,\mathrm{pc}^{-2}$ at $0.1\,\mathrm{kpc}$. The lack of feedback during the formation of the disc galaxy means that we achieve higher than typically observed central dust densities. In addition, the idealized nature of the simulation results in the circumgalactic medium (CGM) of the halo being free from dust because dust is not produced in the CGM but instead is acquired through in-falling satellites.

The evolution of the dust-to-gas ratio radial profile during the formation of the disc galaxy is shown in the right panel of Fig. \ref{fig:ICprops}. The dust initial forms in a very compact region with a radius less than $1\,\mathrm{kpc}$, with a dust-to-gas ratio of $0.05$ at $0.1\,\mathrm{kpc}$ that very rapidly declines with radius. From $0.5\,\mathrm{Gyr}$ to $2.5\,\mathrm{Gyr}$ there is a modest increase in the central dust-to-gas ratio. Over time the radial extent of the dust grows, matching the growth of the stellar disc. At $2.5\,\mathrm{Gyr}$ the dust disc has a radial extent of $8\,\mathrm{kpc}$ and the dust-to-gas ratio decreases from a central value of $0.06$ to $0.001$ at the edge of the disc. This compares to the constant value of $D=0.01$ assumed in previous work.

The key result of the formation of the galaxy is that the optical depth of the halo increases significantly. In Fig \ref{fig:gal_tau} we plot the optical depth integrated to $r_{200}$ as a function of radius for the analytic model, the initial NFW halo assuming $D=0.01$, the halo with a disc galaxy assuming $D=0.01$ and the halo with a disc galaxy in which the formation of dust has been modelled. The analytic and initial NFW halo show excellent agreement with an optical depth $\tau=11.37$ at $0.1\,\mathrm{kpc}$. The formation of the galaxy causes the centre of the halo to contract and the optical depth increases to $\tau=225.41$ at $0.1\,\mathrm{kpc}$. The optical depth declines more steeply with radius and the initial NFW halo has the greater optical depth beyond $5\,\mathrm{kpc}$. Modelling the formation of dust leads to an even greater optical depth in the centre of the halo, reaching $\tau=1044.4$ at $0.1\,\mathrm{kpc}$. However, the optical depth of the halo drops rapidly, due to the steep decline of the dust-to-gas ratio with radius and the lack of dust in the CGM, and it becomes negligible beyond the galactic disc at $8\,\mathrm{kpc}$.

\subsection{Outflow properties}
We now explore the impact of the formation of the galaxy and the assumed dust model has on the properties of the radiatively driven outflow. Four numerical setups are used: an NFW halo with a constant dust-to-gas ratio $D=0.01$ (NFW\_UV47\_m1e5), an NFW halo with a central disc galaxy and constant dust-to-gas ratio $D=0.01$ (Gal\_UV47), an NFW halo with a central disc galaxy and a dust distribution modelled following \citetalias{McKinnon2017} (Gal\_UV47\_McK17) and an NFW halo with a central disc galaxy and a \citetalias{McKinnon2017} dust distribution, but neglecting dust destruction mechanisms (Gal\_UV47\_McK17nd). For those runs where the physical processes of galaxy formation are active, i.e. those with a disc galaxy, these processes continue to operate during the RHD simulations. We have confirmed that this has a negligible impact on the results presented. The formation of the galaxy and modelling the production of dust leads to central optical depths in excess of $1000$ for IR radiation, due to the unrealistically high IR dust opacity we assume. At these optical depths the diffusion speed of IR photons becomes equivalent to, or lower than, the propagation speed of the outflow and they will not efficiently couple momentum to the outflow for simulations with a central galaxy. Therefore, we neglect reprocessing of the UV radiation to IR radiation in this Section and focus on the impact of the central disc galaxy and modelling the distribution of dust grains.

At the centre of each halo is a black hole that isotropically emits UV radiation at a constant bolometric luminosity $L_{\mathrm{AGN}}=10^{47}\,\mathrm{erg}\,\mathrm{s}^{-1}$ throughout the $100\,\mathrm{Myr}$ of the simulation. The black hole is present during the relaxation and the galaxy formation runs, remaining within $0.1\,\mathrm{kpc}$ of the potential minimum at all times. Material is defined as part of the outflow if its halocentric radial velocity is greater than $50\,\mathrm{km}\,\mathrm{s}^{-1}$. 

\begin{figure}
 \includegraphics[width=\columnwidth]{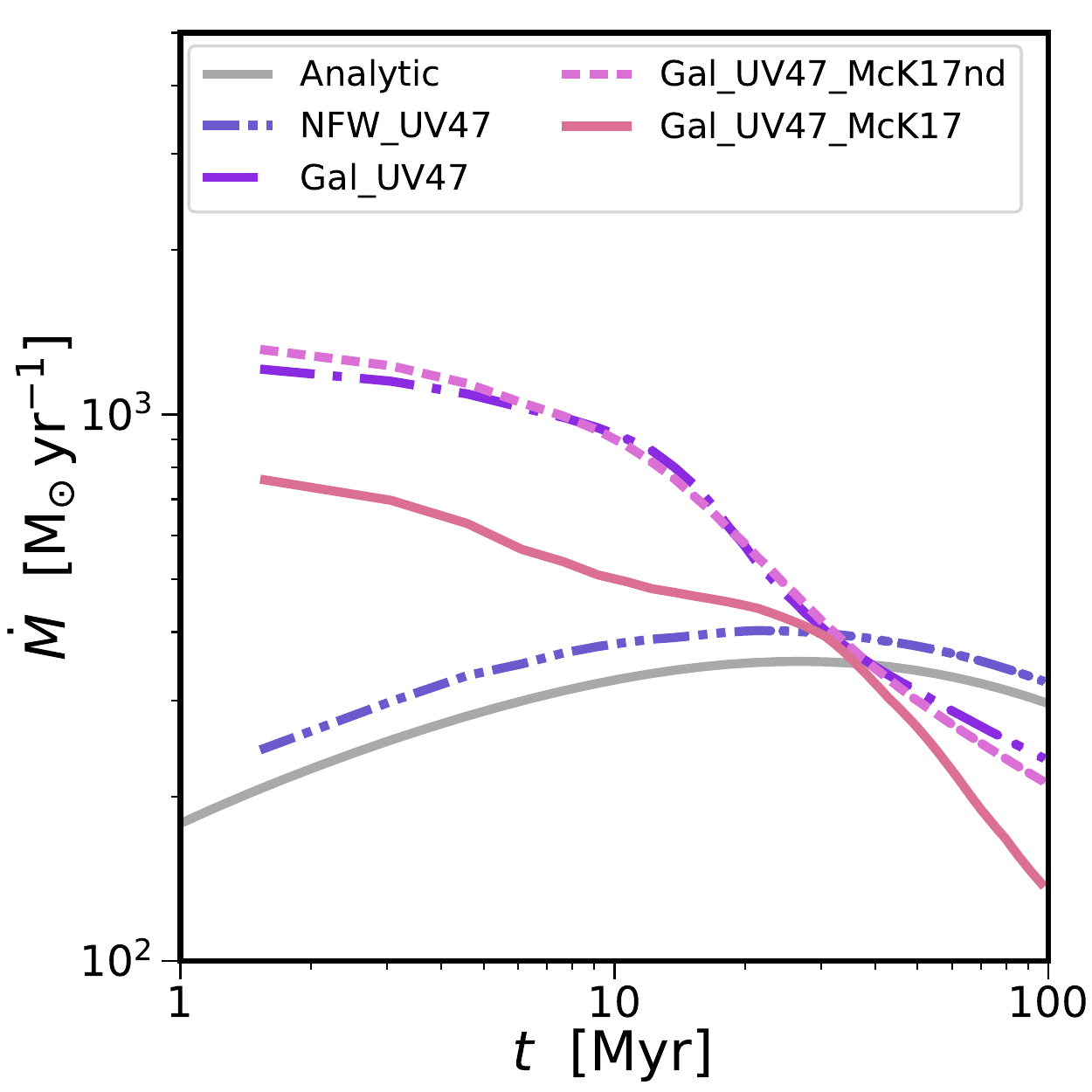}
 \caption{Impact of the assumed dust model on the mass outflow rate as a function of time for a black hole with a bolometric luminosity $L_{\mathrm{AGN}}=10^{47}\,\mathrm{erg}\,\mathrm{s}^{-1}$. We compare the analytic result (grey solid line) to simulated results for an NFW halo with $D=0.01$ (blue dash-dot-dot), an NFW halo with a central disc galaxy and $D=0.01$ (purple dash-dot) and an NFW halo with a central disc galaxy and the dust model of \citet{McKinnon2017} without (pink dashed) and with (red solid) dust destruction mechanisms. The presence of a central galaxy leads to a factor $5$ increase in the initial mass outflow rates relative to the pure NFW halo, due to the increased optical depth of the halo. However, the inclusion of dust destruction mechanisms reduces the mass outflow rate by $40$ per cent relative to the other disc galaxy RHD simulations.}
 \label{fig:Mdot_dust}
\end{figure} 

The mass outflow rate, defined in equation \ref{eq:Mout}, as a function of time for the different setups is shown in Fig. \ref{fig:Mdot_dust}. The relaxed NFW halo is in reasonable agreement with the expected analytic solution, converging at late times. The formation of a disc galaxy (Gal\_UV47), but still assuming $D=0.01$ results in a factor $5$ increase in the initial mass outflow rate, from $243.4\,\mathrm{M}_{\astrosun}\,\mathrm{yr}$ to $1211.6\,\mathrm{M}_{\astrosun}\,\mathrm{yr}$ at $t=1.5\,\mathrm{Myr}$. This is driven by the increased optical depth of the halo due to its contraction during the formation of the galaxy, enabling the radiation to couple more momentum to the gas. However, for $t>30\,\mathrm{Myr}$ the outflow rate drops below that of the NFW halo. The late time mass outflow rate drops for two reasons. First, the outflow has a larger velocity relative to the NFW outflow and reaches a larger radius at a fixed time, which reduces the mass outflow rate due to its $1/r$ radial dependence. In addition, at a larger radius the outflow becomes more diffuse with a lower optical depth, which reduces the efficiency with which radiation can couple momentum to the outflow and the outflow decelerates faster due to the ram pressure of the ambient medium of the halo.

\begin{figure}
 \includegraphics[width=\columnwidth]{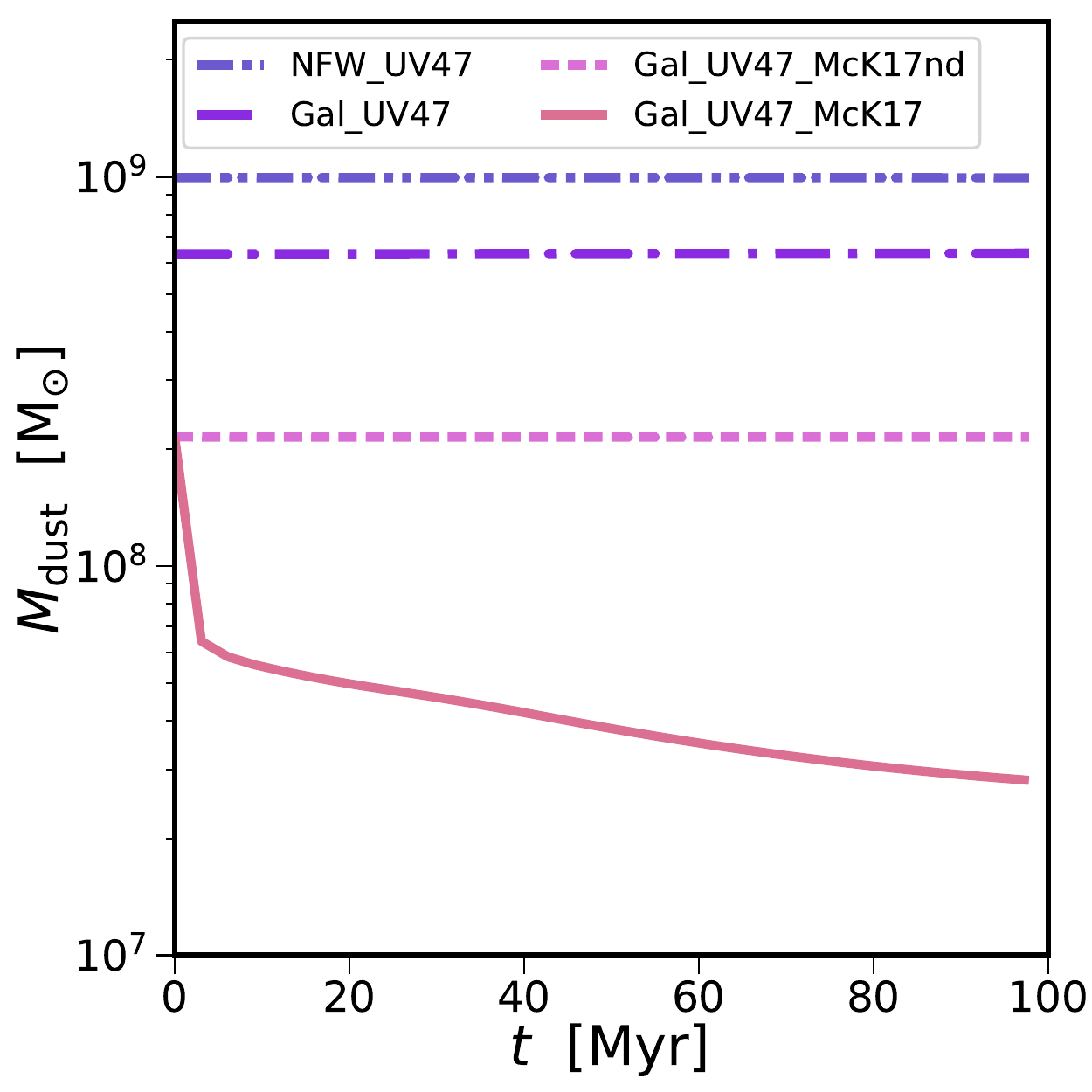}
 \caption{Evolution of the total dust mass within $r_{200}$ as a function of time. Line styles are the same as Fig. \ref{fig:Mdot_dust}. The formation of the disc galaxy removes gas and decreases the total gas mass while modelling the production of dust reduces the dust content of the halo by a factor $5$ relative to the initial NFW halo. However, the inclusion of dust destruction mechanisms leads to a rapid decrease in the dust content of the halo, with $70$ per cent of the dust destroyed after $1.5\,\mathrm{Myr}$ and $87$ per cent removed by $100\,\mathrm{Myr}$.}
 \label{fig:Mdust}
\end{figure}

Neglecting the destruction of dust by shocks and thermal sputtering, the disc galaxy with a modelled dust distribution (Gal\_UV47\_McK17nd) yields a mass outflow rate that is very similar to the constant dust-to-gas ratio galaxy. It has a mass outflow rate of $1316.6\,\mathrm{M}_{\astrosun}\,\mathrm{yr}$ at $t=1.5\,\mathrm{Myr}$ and the mass outflow rate declines strongly with time as the dust is driven away from the black hole. This decline occurs for the same reasons as the Gal\_UV47 simulation. If the destruction of dust grains via shocks and thermal sputtering is included (Gal\_UV47\_McK17) the mass outflow rate at $t=1.5\,\mathrm{Myr}$ is reduced by $42$ per cent to $761.8\,\mathrm{M}_{\astrosun}\,\mathrm{yr}^{-1}$ relative to the Gal\_UV47\_McK17nd simulation. The mass outflow then declines with time, but not as strongly as the runs with a constant dust-to-gas ratio or that neglect destruction mechanisms. At $t=30\,\mathrm{Myr}$ the mass outflow rates all of the simulations with a central galaxy are comparable to each other, at which point the mass outflow rate declines steeply with time for the run with dust destruction. This decline is driven by the lack of dust outside the galaxy, without a supply of dust grains that can be entrained in the outflow the optical depth of the outflow declines with time as dust destruction mechanisms remove dust from the outflow.

The reason for the difference in the mass outflow rates with and without dust destruction becomes apparent when examining the dust mass within $r_{200}$ as a function of time, as shown in Fig. \ref{fig:Mdust}. A constant dust-to-gas ratio leads to a dust mass that is proportional to the gas mass. The reduction in dust mass between the NFW\_UV47\_m1e5 and Gal\_UV47 is due to the conversion of gas to stars during the galaxy's formation. Modelling the production of dust following \citetalias{McKinnon2017} results in a factor $6$ reduction in the total dust mass for simulation Gal\_UV47\_McK17nd, with the galaxy containing $2.14\times10^{8}\,\mathrm{M}_{\astrosun}$ of dust. These three simulation neglect destruction mechanisms and the total dust remains constant for the RHD simulations. However, in Gal\_UV47\_McK17 the destruction of dust grains via shocks and thermal sputtering is included and the dust mass decreases rapidly with time. After $t=3\,\mathrm{Myr}$ the shock heating of the gas by the feedback results in a $70$ per cent reduction from the initial dust mass, mainly due to thermal sputtering. After $100\,\mathrm{Myr}$ the total dust mass within $r_{200}$ is $2.78\times10^{7}\,\mathrm{M}_{\astrosun}$, an $87$ per cent reduction.

\begin{figure}
 \includegraphics[width=\columnwidth]{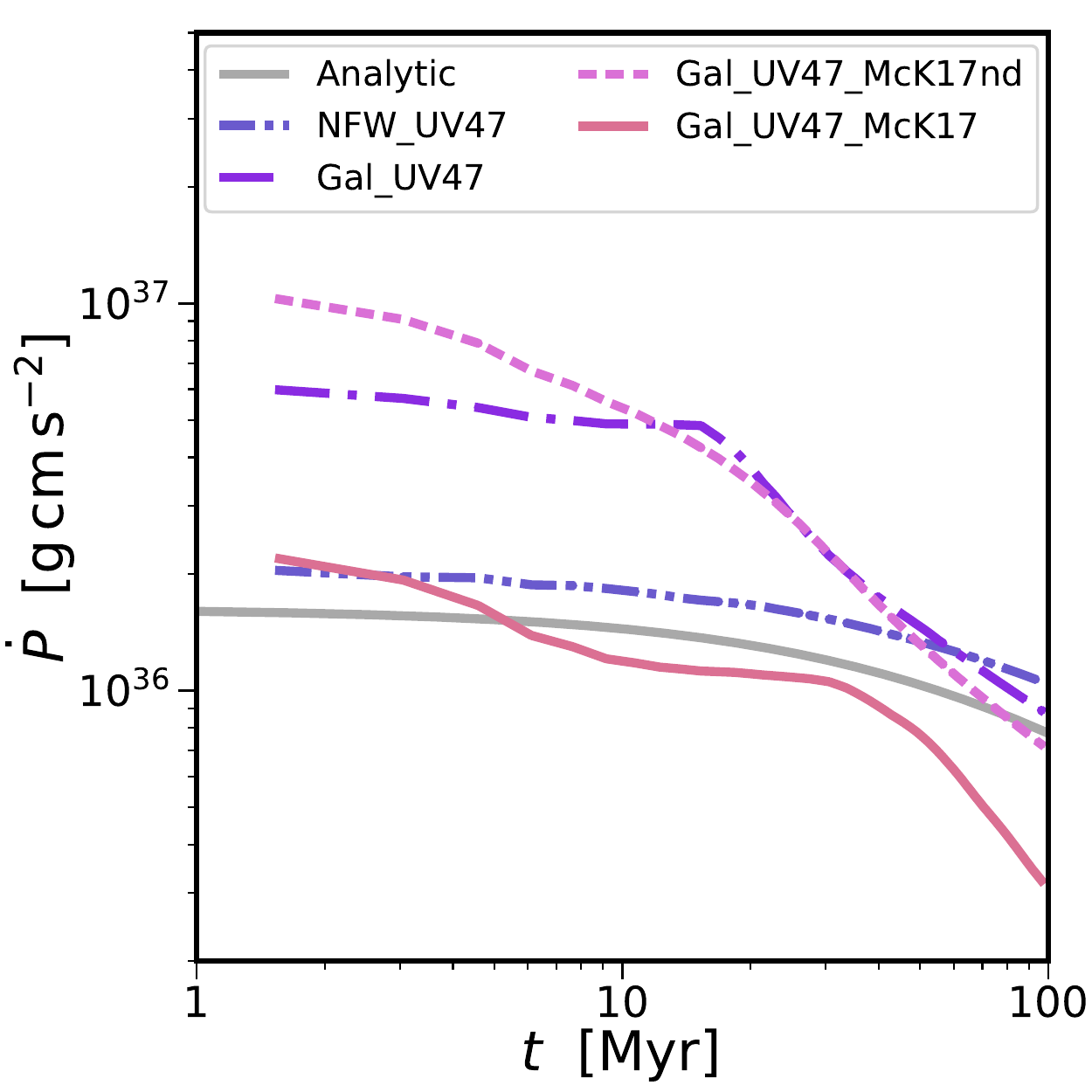}
 \caption{Momentum flux of the outflow as a function of time demonstrating the impact of a central disc galaxy and modelling the production and destruction of dust. Line styles are the same as Fig. \ref{fig:Mdot_dust}. The increased optical depth due to a central galaxy and modelled dust production increases the efficiency of momentum coupling and the generation of stronger shock, increasing the momentum flux of the outflow. However, destruction of dust grains significantly reduces the momentum flux because the radiation cannot couple momentum as efficiently to the gas.}
 \label{fig:Pdot_dust}
\end{figure} 

In Fig. \ref{fig:Pdot_dust} we plot the momentum flux of the outflow as a function of time for the different simulations. The presence of a central disc galaxy leads to a contraction of the halo and increases its optical depth. The radiation then couples momentum to the gas more efficiently and drives a stronger shock. This increase the moment flux of the outflow relative to the pure NFW halo, yielding a value of 
$5.98\times10^{36}\,\mathrm{g}\,\mathrm{cm}\,\mathrm{s}^{-1}$ at $t=1.5\,\mathrm{Myr}$ for the Gal\_UV47 simulation. This value remains relatively flat until $t=20\,\mathrm{Myr}$ as the outflow sweeps up the high-density material of the galactic disc, before declining as the outflow expands into lower density halo. If the distribution of dust is modelled the momentum flux increases by $60$ per cent relative to the constant dust-to-gas ratio simulation. This is because modelling the formation further increases the optical depth of the halo. The stronger shock produced in the Gal\_UV47\_McK17nd produces a higher velocity outflow, sweeping up the disc material more rapidly so that the momentum flux declines with time rather than remaining flat for a period of time. If dust destruction mechanisms are included the peak momentum flux achieved is $2.20\times10^{36}\,\mathrm{g}\,\mathrm{cm}\,\mathrm{s}^{-1}$, which produces a momentum ratio $\zeta=0.65$ that is below typically observed values. The momentum flux remains relatively flat as the outflow sweeps up the material in the galactic disc before it then declines with time. The reduced momentum flux is driven by the rapid destruction of the dust at early times, which significantly reduces the optical depth of the halo, the momentum the radiation can couple to the gas and the velocity of the outflow.

\begin{figure}
 \includegraphics[width=\columnwidth]{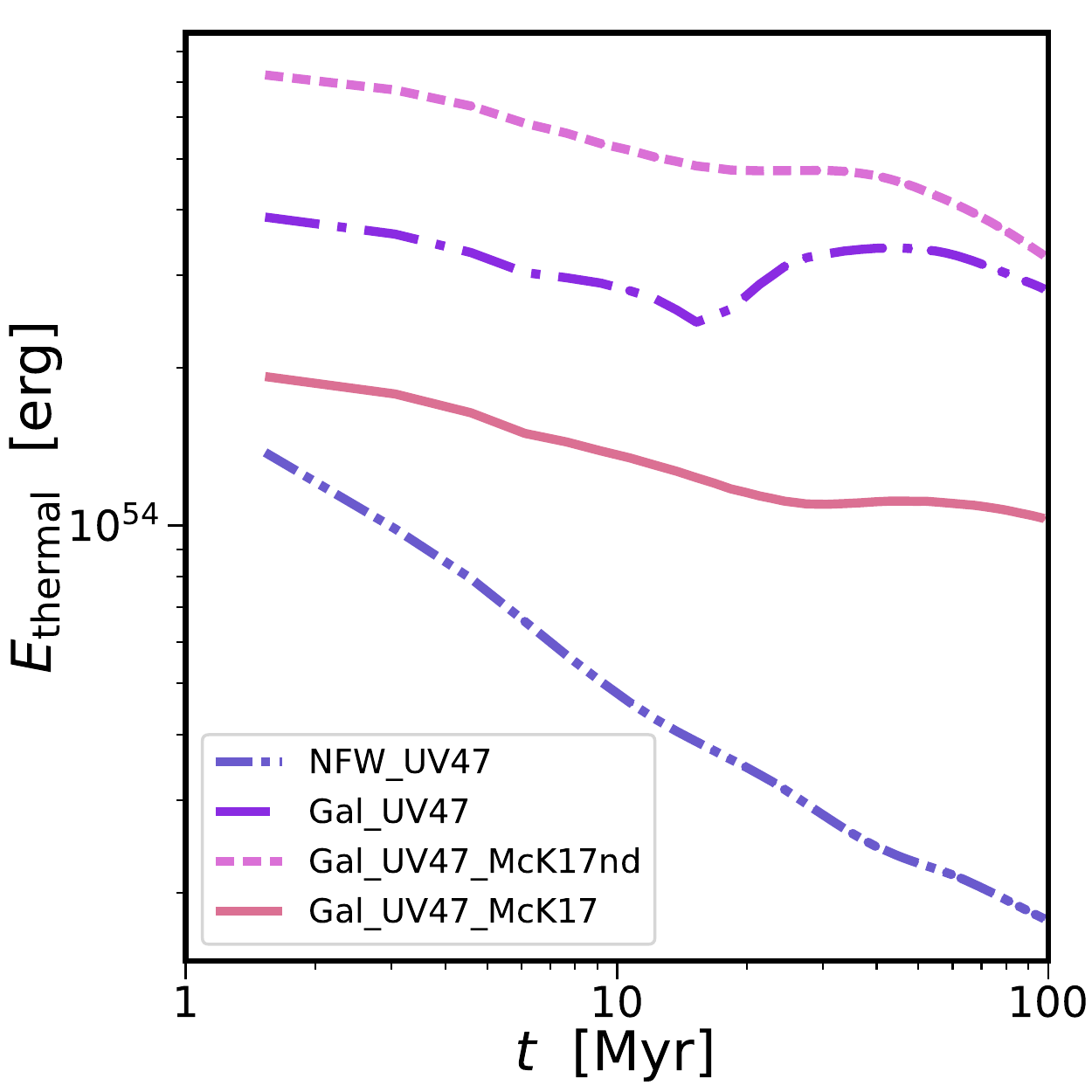}
 \caption{Mass-weighted thermal energy of the outflow as a function of time. Line styles are the same as Fig. \ref{fig:Mdot_dust}. The increased optical depth of the halo leads to the generation of a stronger shock because the radiation couples momentum to the gas more efficiently. This stronger initial shock heats the outflow material and increases its thermal energy content relative to the pure NFW halo simulation.}
 \label{fig:Ethm}
\end{figure}

At $t=1.5\,\mathrm{Myr}$ the Gal\_UV47 and Gal\_UV47\_McK17nd simulations have momentum ratios of $\zeta=1.79$ and $3.09$, respectively, in agreement with typical observed values of $\zeta=1-30$ \citep{Cicone2014,Zakamska2014,Carniani2015,Fiore2017} and larger than $\zeta=1$ maximum for simulations without IR radiation. Therefore, the outflows must adiabatically expand and do 'PdV' work to boost the momentum ratio above unity. In Fig. \ref{fig:Ethm} we plot the mass-weighted thermal energy of the outflow as a function of time. The higher optical depth of simulations with a central disc galaxy and a modelled dust distribution lead to a $260$ and $489$ per cent increase in the thermal energy of the outflow at $t=1.5\,\mathrm{Myr}$ for the Gal\_UV47 and Gal\_UV47\_McK17nd simulations, respectively, relative to NFW\_UV47\_m1e5 simulation. The thermal energy of outflow in the Gal\_UV47\_McK17 is larger than the pure NFW simulation, however, the rapid destruction of the dust reduces the efficiency with which radiation can couple momentum to the gas, so the momentum ratio of the outflow is less than unity before $t=1.5\,\mathrm{Myr}$, the first output of the simulation.

\begin{figure}
 \includegraphics[width=\columnwidth]{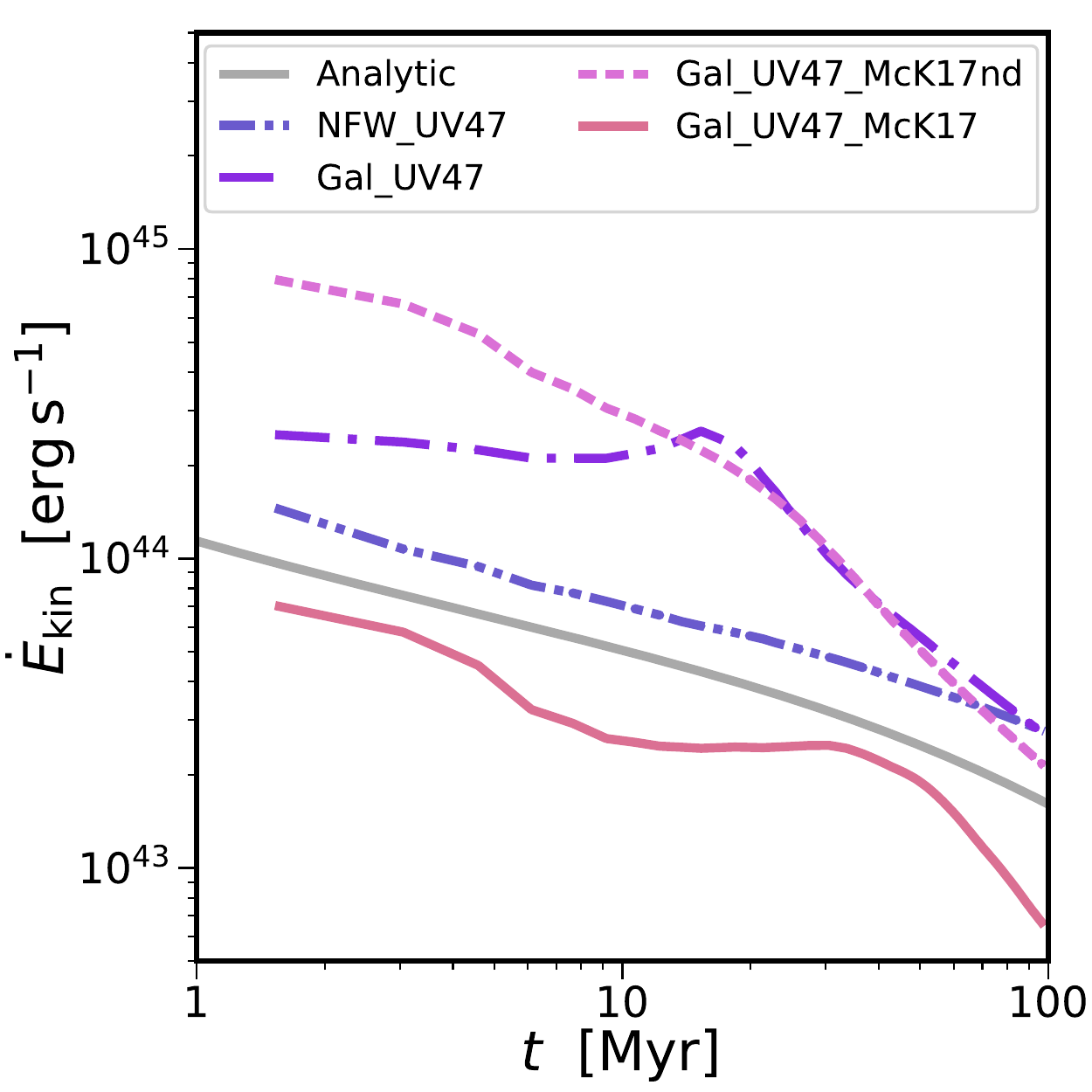}
 \caption{Kinetic power of the outflow as a function of time demonstrating the impact of a central disc galaxy and modelling the production and destruction of dust. Line styles are the same as Fig. \ref{fig:Mdot_dust}. The increased coupling efficiency of the radiation and the increased work done by the expansion of the hotter outflow material leads to an increased kinetic power. However, dust destruction mechanisms reduce the coupling efficiency of the radiation and kinetic power of the outflow.}
 \label{fig:Ekin_dust}
\end{figure}

The kinetic power of the outflow produces a very similar result. In Fig. \ref{fig:Ekin_dust} we plot the kinetic power as a function of time for the 4 numerical setups. The formation of the disc galaxy and assuming a constant dust-to-gas ratio leads to kinetic power that increases marginally with time to $t=20\,\mathrm{Myr}$ as the outflow sweeps out the gas in the galactic disc, before declining with time once the outflow begins to expand into the halo gas. A disc galaxy with a modelled dust distribution results in an outflow whose kinetic power decreases with time because it rapidly sweeps up the disc material and dust is only present in the disc of the galaxy. At $t=1.5\,\mathrm{Myr}$ these simulations achieve a kinetic power of $2.51\times10^{44}\,\mathrm{erg}\,\mathrm{s}^{-1}$ and $7.97\times10^{44}\,\mathrm{erg}\,\mathrm{s}^{-1}$, respectively. This yields energy ratios of $0.25$ and $0.97$ per cent, respectively, in agreement with observed values \citep{Cicone2014,Zakamska2014,Carniani2015,Zakamska2016,Fiore2017}. However, the incorporation of dust destruction processes yields a reduced mass outflow travelling at a decreased radial velocity, and the kinetic power of the outflow at $t=1.5\,\mathrm{Myr}$ is $7.04\times10^{43}\,\mathrm{erg}\,\mathrm{s}^{-1}$. This produces an energy ratio of $0.07$ per cent, well below observed values because the destruction of the dust reduces the efficiency with which radiation can couple momentum to the outflow.

\begin{figure}
 \includegraphics[width=\columnwidth]{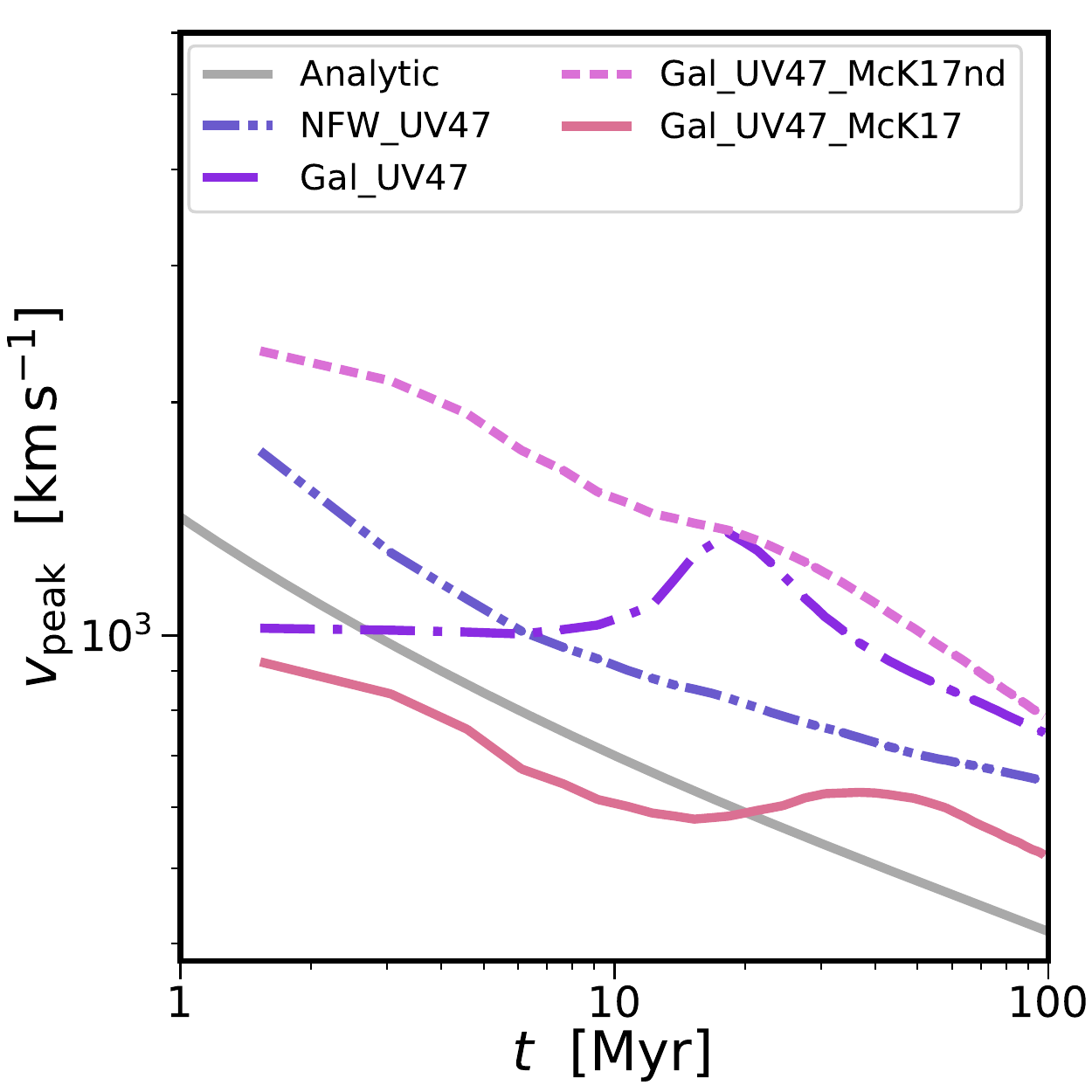}
 \caption{Peak velocity of the outflow as a function of time for the pure NFW, a central disc galaxy and a modelled dust distribution. Line styles are the same as Fig. \ref{fig:Mdot_dust}. A central disc galaxy and modelling dust production results in a higher peak velocity due to the increased optical depth and stronger result outflow shock. However, the destruction of the dust reduces the peak velocity by more the than a factor of $2$.} 
 \label{fig:Vpeak_dust}
\end{figure}

Finally, we explore the impact of the formation of a galaxy and the assumed dust model on the peak velocity of the outflow. Fig. \ref{fig:Vpeak_dust} shows the peak velocity as a function of time. The peak velocity of the simulated outflow is taken as the $95^{\mathrm{th}}$ percentile of the velocity distribution of material classified as being part of the outflow. The formation of the galaxy with a constant dust-to-gas ratio leads to peak velocity that is initially flat as the outflow sweeps up the gas in the galactic disc, with the peak outflow velocity of $1355\,\mathrm{km}\,\mathrm{s}^{-1}$ at $t=18\,\mathrm{Myr}$. After this time the peak velocity of the outflow decays with time as the outflow expands into the low-density halo gas and the optical depth of the outflow reduces. Modelling the formation of the dust leads to a higher peak velocity at early times that constantly reduces with time, achieving a peak velocity of $2330\,\mathrm{km}\,\mathrm{s}^{-1}$ at $t=1.5\,\mathrm{Myr}$. The increased velocity of the outflow is driven by the increased optical depth of the halo, which enables the radiation to deposit more momentum into the gas. However, including dust destruction processes leads to a dramatically reduced outflow velocity. With significantly less dust, the optical depth of the halo is reduced and the initial outflow achieves a velocity of $925\,\mathrm{km}\,\mathrm{s}^{-1}$ at $t=1.5\,\mathrm{Myr}$. The peak velocity of the outflow is a result of the competition between the rate at which dust is destroyed and the rate at which the outflow sweeps up fresh dust from the galactic disc. As the outflow reaches the edge of the disc the velocity briefly rises as it sweeps up sufficient dust to increase its optical depth. This dust is then destroyed and the outflow expands into the halo gas that is devoid of dust and the velocity of the outflow declines.

The formation of a disc galaxy significantly increases the optical depth of the halo, which in turn significantly increases the peak mass outflow rate, momentum flux, kinetic power and peak velocity of the radiatively driven outflow. Modelling the production of dust during the formation of the galaxy does not lead to a significant difference in the mass outflow rate. The increased velocity of the initial outflow increases its momentum flux and kinetic power, yielding momentum and kinetic ratios in broad agreement with observed outflows. However, when dust destruction processes are included the dust in the halo is very quickly destroyed by thermal sputtering. This reduces the optical depth of the halo and the outflow measures are significantly reduced.

\section{Discussion}
\label{sec:lims}
In this work we have extended previous analytic and numerical work on radiative AGN feedback to explore the impact of a central galaxy and eliminated the need to assume a constant dust-to-gas ratio by modelling the formation of dust. These extensions have a profound impact on the outflow generated by radiative feedback. Driven by the optical depth of the halo, modelling the formation of a galaxy and the dust grains within it yields better agreement between the properties of simulated and observed outflows, but the incorporation of dust destruction mechanisms leads to a rapid decrease in the dust content of the halo and the simulated outflow properties are lower than typically observed. However, there are physical and numerical limitations of our method that potentially impact the presented results and we discuss these now.

\subsection{Isolated NFW halo}
An outflow generated in an isolated halo must overcome the ram pressure confinement of the ambient medium of the halo as it propagates outwards (see Section \ref{sec:sim_anlyt}). However, in a cosmological setting material infalling onto the halo will increase the ram pressure experienced by the outflow. This additional pressure will impact the evolution of the outflow properties with time, increasing the proportion of the outflow material that stalls in the outskirts of the halo and reducing the outflow properties at a fixed time relative to the isolated halo setup.

The combination of stalled outflow material and inflowing material will form the circumgalactic medium of the halo, a component of the halo that is poorly modelled in our numerical setup. Cold material present in the CGM, whether accreted cold or outflow material that has cooled \citep[e.g.][]{Zubovas2014}, is potentially important for the dynamics of outflows generated by radiative feedback. In this work the formation of dust occurs only in the disc of the galaxy. However, cold material in the CGM can harbour dust grains. As the outflow propagates through the CGM it will sweep up this material, increasing its dust content, its optical depth and the momentum the radiation can couple to the outflowing material. CGM dust is potentially important when dust destruction mechanisms, like thermal sputtering, rapidly destroy dust entrained in the outflow. Improved modelling of the CGM potentially could have a significant impact on the time evolution of the simulated outflow properties.

\subsection{Galaxy formation model}
The scales and uncertain physics involved in galaxy formation make the use of subgrid models a necessity in numerical simulations. In this study we make use of an effective equation of state for the dense interstellar medium (ISM) that is star-forming \citep{Springel2003}. This models the multiphase, quasi-equilibrium structure of the ISM as a single effective-temperature gas. The multiphase nature of the ISM is important for the behaviour of radiatively driven outflows. Dense, cold gas in the ISM will shield the dust within it when entrained in the outflow, increasing the optical depth of the outflow and reducing the rate at which dust grains are destroyed. In addition, previous studies have shown that the outflow takes the path of least resistance around cold dense gas \citep{Costa2014,Bourne2014,Gabor2014,Bieri2017}. The result is that the amount of cold gas that becomes entrained within the outflow will depend on its distribution within the ISM and the CGM, impacting the efficiency with which radiation couples to the outflow. Therefore, a model of the ISM that is able to capture its multiphase nature may significantly impact the properties of outflows generated by radiative feedback.

To focus on the impact of radiative feedback from the central black hole we have neglected feedback from stars in this work. The inclusion of stellar radiation has been shown to suppress the star formation rate of the host galaxy via photo-heating of the gas \citep[e.g.][Kannan et al. in prep.]{Rosdahl2015,Ishiki2017}. Photo-heating reduces the density of the ISM, decreasing its optical depth and the efficiency with which radiation can couple to the gas. The combination of radiation and supernovae feedback from stars acts to preprocess the ISM, creating low-density channels through which radiation can escape. This potentially enhances the properties of outflows driven by radiative AGN feedback, as the outflow escapes through these channels and the ram pressure confinement of the ambient medium is reduced. However, it also lowers the optical depth of the halo, reducing the efficiency with which radiation can couple to the gas and reducing observable outflow characteristics. Therefore, it is critical to study the individual and combined effects of different stellar feedback mechanisms on radiatively driven outflows.

Finally, the treatment of the black hole was deliberately kept simple in this study to compare with analytic work. The black hole does not accrete material and it releases radiation at a constant luminosity in a single band, independent of the physical conditions around it. Although accretion physics is highly uncertain, to some degree the luminosity of the AGN should be related to the accretion rate of the black hole and this should depend on the local properties of the gas around the black hole. This will lead to self-regulation of the feedback when the density of material around the black hole is high the accretion rate will increase and AGN luminosity increases. The radiation couples more efficiently to the surrounding gas because its density, and resulting optical depth, is enhanced. The outflow produced by the feedback decreases the density around the black hole, reducing the accretion rate, the AGN luminosity and the coupling efficiency of the radiation to the point where it is no longer able to drive an outflow. The material then begins to cool and condense around the black hole, increasing the local density and beginning the cycle again. A more thorough treatment of the black hole will likely have a significant impact on the outflows it produces.

\subsection{Dust physics}
Previous analytic and numerical work assumed that the dust-to-gas ratio was fixed at a constant value throughout the halo, an assumption that we explored the impact of by modelling the production of dust during the formation of the galaxy. However, a limitation of our approach is that the dust is treated as scalar quantity with a single grain size that is advected with the movement of the gas, i.e. it is perfectly hydrodynamically coupled. In reality, the coupling between dust grains and gas depends on the size distribution of the dust grains \citep[e.g.][]{Hopkins2016b}. Any decoupling of the dust grains and the gas will limit the momentum the radiation can impart and the properties of the generated outflows. An improved approach would be to model the dust as a separate fluid \citep[e.g.][]{Bekki2015,Booth2015,McKinnon2018} and then assess how coupled the dust grains and gas remain in the presence of the strong shocks driven by radiative AGN feedback.

In this work we modelled the production, growth and destruction of dust grains. However, the interaction of the radiation with the dust grains can be further developed. First, the Rosseland mean opacity of the dust is assumed to be fixed for all temperatures, with an unrealistically high value assumed for IR radiation (to match previous work and demonstrate the momentum boost it can provide). In reality the opacity of varies as a function of temperature \citep{LiDraine2001,Semenov2003}, with the opacity for IR photons scaling very strongly at low $(T<150\,\mathrm{K})$ temperatures. This would alter the optical depth of the halo and the resulting efficiency with which radiation can couple momentum to the gas. Finally, photoevaporation of the dust is potentially an important mechanism for dust destruction that has been neglected, though it would occur on spatial scales that are not resolved by these simulations.

\section{Conclusions}
\label{sec:concs}
The impact of radiation in the Universe is only just beginning to be understood. In this work we have explored radiation pressure on dust grains as a channel of AGN feedback. Adopting the simplified setup of an isolated NFW halo we compared the propagation of a radiatively driven outflow against the propagation expected from the analytic solution and compared the properties of the outflow to typical values of observed outflows. We then examined the impact of forming a central disc galaxy in the halo and modelling the production and destruction of dust grains, rather than assuming a fixed dust-to-gas ratio. Our conclusions are
\begin{itemize}
 \item The propagation of the simulated shock front is in excellent agreement with the expected analytic solution (Fig. \ref{fig:UVonly}). For the lower luminosity AGN the confining pressure of the ambient medium, which is neglected in the analytic solution, begins to impact the propagation of the shock at late times. Due to the improved convergence properties of the \textsc{Arepo-RT} code, we find that the propagation of the shock is independent of numerical resolution (Fig. \ref{fig:Con_mass}). If the re-processing of absorbed UV radiation into IR radiation is included then the simulations yield the expected momentum boost (Fig. \ref{fig:UV+IR}). The propagation of the simulated shock front is unchanged for any reduced speed of light approximation where the propagation of radiation through the simulation domain is significantly faster than the velocity of the shock front, e.g. $\tilde{c}\geq0.01c$ (Fig. \ref{fig:Con_clight})
 \item Exploring the properties of the generated outflow we found the simulation with isotropic injection is in reasonable agreement with the expected analytic solution. Discrepancies arise because of the finite resolution of the simulation, compared to the thin shell approximation of the analytic solution, and the confining ram pressure of the ambient medium of the halo. As the radiation injection region becomes more collimated the mass outflow rate (Fig. \ref{fig:Mdot}) and momentum flux (Fig. \ref{fig:Pdot}) of the outflow is reduced. In contrast the kinetic power (Fig. \ref{fig:Ekin}) and peak velocity (Fig. \ref{fig:Vpeak}) of the outflow increases. These results are driven by the radiation coupling momentum to a smaller amount of mass, but accelerating it large velocities.
 \item Regardless of the how collimated the radiation injection is we find that the simulated outflows in the simplified NFW setup are unable to reproduce the typical values of observed ionized outflows. The simulated outflows achieve a peak momentum ratio $\zeta\approx0.3$ and peak energy ratio $\epsilon_{\mathrm{k}}\approx0.25$ compared to $\zeta=1-30$ and $\epsilon_{\mathrm{k}}=0.3-3$ for observed outflows. Collimation of the radiation does not lead to a significantly stronger shock and work done by the adiabatic expansion of the outflow remains subdominant.
 \item Allowing the gas in the NFW halo to radiatively cool results in the formation of a Milky Way-\textit{like} galaxy (Fig. \ref{fig:ICprops}). The formation of the galaxy leads to a factor $15$ increase in the central dust surface density, assuming a dust-to-gas ratio $D=0.01$. Modelling the production of dust during the formation of the galaxy further increases the central dust surface density, with a central dust-to-gas ratio $D=0.06$. The key result is that the optical depth of the halo increases by two orders of magnitude relative to the initial NFW halo (Fig. \ref{fig:gal_tau}).
 \item The properties of outflows generated in a halo with a central disc galaxy and a constant dust-to-gas ratio are in better agreement with typically observed values, with a factor $\sim5$ increase in the mass outflow rate (Fig. \ref{fig:Mdot_dust}), momentum flux (Fig. \ref{fig:Pdot_dust}), kinetic power (Fig. \ref{fig:Ekin_dust}) and peak velocity (Fig. \ref{fig:Vpeak_dust}). Modelling the production of dust leads to a similar mass outflow rate, but increases in the momentum flux, kinetic power and peak velocity. These values increase because the radiation couples more efficiently to the gas due to the increased optical depth. This drives a stronger shock, which increases the thermal energy of the outflow and the amount of work done by its expansion. The combination of these two effects leads to momentum ratios above unity.
 \item However, once dust destruction mechanisms are included the outflow properties are reduced and lower than typically observed. This is driven by the rapid destruction of dust grains when they are swept up by the outflow, with $70$ per cent of the initial dust mass destroyed within $1\,\mathrm{Myr}$ of the AGN switching on (Fig. \ref{fig:Mdust}). The shock generated by the AGN feedback heats the outflow to $>10^{7}\,\mathrm{K}$ and thermal sputtering of the dust grains leads to their rapid depletion.
\end{itemize}
The results presented in this work extend previous analytic and numerical work regarding radiation pressure on dust as a channel of AGN feedback. We modelled a more realistic halo with a central galaxy and removed the need to assume a constant dust to gas ratio. Despite the limitations of this work, we find that radiation pressure on dust can launch outflows for AGN with sufficient bolometric luminosities, even when the vast majority of the dust grains are destroyed in the outflow. To further investigate radiative AGN feedback requires a more realistic treatment of the host galaxy, its ISM and the physics relevant to dust grain formation and destruction.

\section*{Acknowledgements}
We thank Volker Springel for making \textsc{Arepo} available. The simulations were performed on the MKI-Harvard Odyssey cluster and the Comet supercomputer at the San Diego Supercomputer Center as part of XSEDE project AST180025. RK acknowledges support from NASA  through Einstein Postdoctoral Fellowship grant  number PF7-180163 awarded by the \textit{Chandra} X-ray Center, which is operated by the Smithsonian Astrophysical Observatory for NASA under contract NAS8-03060. MV acknowledges support through an MIT RSC award, a Kavli Research Investment Fund, NASA ATP grant NNX17AG29G, and NSF grants AST-1814053 and AST-1814259.

\bibliographystyle{mnras}
\bibliography{main}

\bsp	
\label{lastpage}
\end{document}